\documentclass[aps,prx,twocolumn,superscriptaddress,longbibliography]{revtex4-2}

\usepackage{tikz}
\usetikzlibrary{quantikz}
\usepackage{tikz-cd}
\tikzcdset{scale cd/.style={every label/.append style={scale=#1},
    cells={nodes={scale=#1}}}}
\usepackage{lipsum}
\usepackage{adjustbox}
\usepackage{longtable}
\usepackage{colortbl}
\usetikzlibrary{arrows}
\usepackage{ctable}
\tikzset{
operator/.append style={fill=purple!20},
my label/.append style={above right,xshift=0.3cm},
phase label/.append style={label position=above}
}
\usepackage[colorlinks=true,breaklinks=true,linkcolor=magenta,citecolor=magenta,urlcolor=magenta]{hyperref}
\usepackage{url}

\usepackage[ruled,vlined]{algorithm2e}
\usepackage{amsmath,amsfonts,bbm}
\usepackage[colorlinks=true]{hyperref}
\usepackage{graphicx}
\usepackage{amsmath}
\usepackage{listings}
\usepackage{parskip}
\usepackage{braket}
\usepackage{makecell}
\usepackage{tabularx}
\usepackage{array}
\usepackage{float}
\usepackage{placeins}
\usepackage{fancyhdr}
\usepackage{booktabs}
\newcommand\tikznode[3][]%
   {\tikz[remember picture,baseline=(#2.base)]
      \node[minimum size=0pt,inner sep=0pt,#1](#2){#3};%
   }
\tikzset{>=stealth}

\newcommand*{\gateStyle}[1]{{\textsf{\itshape #1}}}
\newcommand*{\hGate}{\gateStyle{H}}
\newcommand*{\xGate}{\gateStyle{X}}
\newcommand*{\sGate}{\gateStyle{S}}
\newcommand*{\tGate}{\gateStyle{T}}
\newcommand*{\yGate}{\gateStyle{Y}}
\newcommand*{\zGate}{\gateStyle{Z}}
\newrobustcmd{\B}{\bfseries}

\newcommand*{\circuitH}{\gate[style={fill=teal!20},label style=black]{\textnormal{\hGate{}}}}
\newcommand*{\circuitX}{\gate[style={fill=blue!20},label style=black]{\textnormal{\xGate}}}
\newcommand*{\circuitS}{\gate[style={fill=orange!20},label style=black]{\textnormal{\sGate}}}
\newcommand*{\circuitT}{\gate[style={fill=orange!20},label style=black]{\textnormal{\tGate}}}

\newcommand*{\circuitY}{\gate[style={fill=yellow!20},label style=black]{\textnormal{\yGate}}}

\newcommand*{\circuitZ}{\gate[style={fill=green!20},label style=black]{\textnormal{\zGate}}}

\begin{document}

\title{Quanto: Optimizing Quantum Circuits with Automatic Generation of Circuit Identities}  

\author{Jessica Pointing} 
\affiliation{Department of Physics, University of Oxford, Oxford, OX1 3PU, United Kingdom}
\affiliation{Department of Computer Science, Stanford University, Stanford, California, 94305, United States}

\author{Oded Padon} 
\affiliation{VMware Research, Palo Alto, California, 94304, United States}
\affiliation{Department of Computer Science, Stanford University, Stanford, California, 94305, United States}

\author{Zhihao Jia} 
\affiliation{Department of Computer Science, Stanford University, Stanford, California, 94305, United States}
\affiliation{Department of Computer Science, Carnegie Mellon University, Pittsburgh, Pennsylvania, 15213, United States}

\author{Henry Ma} 
\affiliation{Department of Computer Science, University of California, Los Angeles, California, 90095, United States}

\author{Auguste Hirth} 
\affiliation{Department of Computer Science, University of California, Los Angeles, California, 90095, United States}


\author{Jens Palsberg} 
\affiliation{Department of Computer Science, University of California, Los Angeles, California, 90095, United States}

\author{Alex Aiken} 
\affiliation{Department of Computer Science, Stanford University, Stanford, California, 94305, United States}

\begin{abstract}
Existing quantum compilers focus on mapping a logical quantum circuit to a quantum device and its native quantum gates. Only simple circuit identities are used to optimize the quantum circuit during the compilation process. This approach misses more complex circuit identities, which could be used to optimize the quantum circuit further.

We propose Quanto, the first quantum optimizer that automatically generates circuit identities. Quanto takes as input a gate set and generates provably correct circuit identities for the gate set.

Quanto's automatic generation of circuit identities includes single-qubit and two-qubit gates, which leads to a new database of circuit identities, some of which are novel to the best of our knowledge. In addition to the generation of new circuit identities, Quanto's optimizer applies such circuit identities to quantum circuits and finds optimized quantum circuits that have not been discovered by other quantum compilers, including IBM Qiskit and Cambridge Quantum Computing Tket. Quanto's database of circuit identities could be applied to improve existing quantum compilers and Quanto can be used to generate identity databases for new gate sets.
\end{abstract}

\maketitle

\section{Introduction}
Quantum programs are expressed as circuits of primitive gates; the specific {\em gate set} available varies depending on the quantum computer. Compilers are used to optimize the circuit in an effort to reduce the execution time, the noise that naturally arises in quantum computations \cite{2019nisq}, or both.
\\ \\
A quantum algorithm is implemented on a quantum computer with a quantum circuit, which consists of a sequence of quantum gates. A quantum gate applies a transformation to a qubit and can be represented as a matrix. Generally, the larger the circuit depth (length of the longest path from the input to the output), the noisier the computation, leading to more errors in the results. Because of the possible circuit of incorrect results and the inherent statistical nature of quantum computation, quantum programs must be executed many times to obtain an accurate picture of their results. Therefore, reducing the depth of a quantum circuit not only makes a quantum program faster, but also reduces the potential for noise and reduces the number of program evaluations required.  Thus, finding equivalent but smaller depth circuits is a core problem in the implementation of quantum programs.
\\ \\
Quantum hardware imposes many constraints on quantum compilers, including the layout and connectivity of the qubits. Such aspects are addressed by the compilers built by the hardware vendor (such as IBM or Google). Less attention has been given to optimizing the logical circuit itself -- finding circuit identities for a given gate set. 

Circuit identities have been found manually \cite{2011equivalentquantumcircuits} and automatically for single-qubit gates \cite{2003quantumcircuitidentities}, but as far as we know there is no system today that automatically finds identities for full gate sets and no compiler that optimizes quantum circuits based on such circuit identities.

\begin{figure}
\includegraphics[width=0.8\columnwidth]{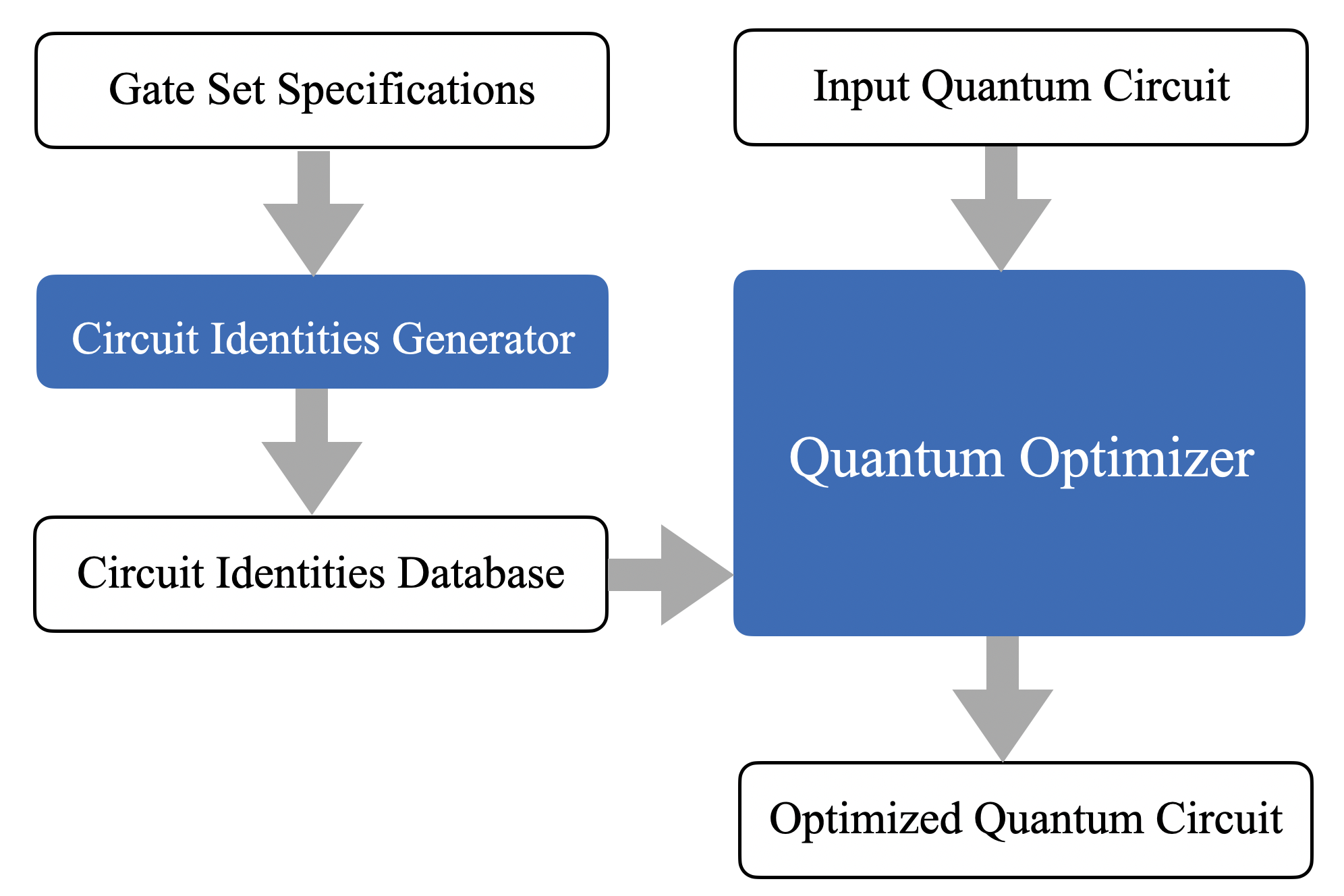}
\caption{Overview of Quanto}
\label{fig:overview_of_quanto}
\end{figure}

We present Quanto, the first quantum circuit optimizer that \textit{automatically generates} circuit identities. Figure \ref{fig:overview_of_quanto} shows an overview of Quanto. In a  first, offline phase, Quanto generates a database of circuit identities. The Quanto compiler takes a quantum circuit and identity database as inputs and uses a cost-based search algorithm to replace the quantum circuit with an optimized quantum circuit. 

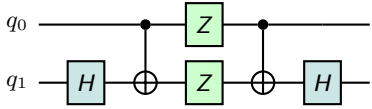
\begin{figure}
    \adjustbox{scale=1}{%
     \begin{tikzcd}
        \lstick{$q_0$} & \qw & \ctrl{1} & \circuitZ & \ctrl{1} & \qw & \qw \\
        \lstick{$q_1$} & \circuitH & \targ{} & \circuitZ & \targ{} & \circuitH & \qw \\
    \end{tikzcd}
    }
\caption{Example of a quantum circuit that can be optimized with Quanto}
\label{fig:example}
\end{figure}

Figure \ref{fig:example} shows an example of a quantum circuit with two qubits and a depth of five. The circuit contains the two-qubit gate $CX$ and the single-qubit gates $H$ and $Z$. The single-qubit gates are represented by $2 \times 2$ unitary complex matrices and the two-qubit gates are represented by $4 \times 4$ unitary complex matrices. This quantum circuit is a \textit{logical quantum circuit}, which is a quantum circuit defined in terms of unitary matrices and is agnostic to the quantum hardware.

If we wish to run this logical quantum circuit on the quantum hardware, we would have to compile the quantum circuit into the hardware's \textit{native quantum gates}. These are the set of quantum gates that have been implemented on particular quantum hardware. For example, IBM's native quantum gate set was \footnote{IBM recently changed their native gate set to \{CX, ID, RZ, SX, X\}} \{U1,U2,U3,CX\} where
\begin{flalign*}
 & U1(\lambda) = \begin{pmatrix}
    1 & 0 \\ 
    0 & e^{i\lambda}
\end{pmatrix} \\
 & U2(\psi,\lambda) = \frac{1}{\sqrt{2}}\begin{pmatrix}
    1 & -e^{i\lambda} \\ 
    e^{i\psi} & e^{i(\lambda+\psi)}
\end{pmatrix} \\
 & U3(\theta,\psi,\lambda) = \begin{pmatrix}
    \cos(\frac{\theta}{2}) & -e^{i\lambda} \sin(\frac{\theta}{2}) \\
    e^{i\psi} \sin(\frac{\theta}{2}) & e^{i(\lambda + \psi)} \cos(\frac{\theta}{2})
\end{pmatrix} \\
 & CX = \begin{pmatrix}
    1 & 0 & 0 & 0\\ 
    0 & 1 & 0 & 0\\
    0 & 0 & 0 & 1\\
    0 & 0 & 1 & 0
\end{pmatrix}
\end{flalign*}

In this example, the \textit{source gates} are the gates present in the logical quantum circuit \{X, H, CZ\} and the \textit{target gates} are the native quantum gates \{U1, U2, U3, CX\}. Therefore, in order to run this logical quantum circuit on the quantum hardware, we must find circuit identities for the source gates in terms of the target gates. In this case $Z = U1(\pi)$ and $H = U2(\pi,0)$.

Another step for running a logical quantum circuit on the hardware is to map the qubit and gate connections to those of the physical constraints of the hardware. Most compilers focus on these two aspects: (1) mapping source gates to target gates and (2) mapping the quantum circuit to the hardware constraints. Then, generally, a set of simple manual rules are used to optimize the logical quantum circuit itself. For example, voqc (Verified Optimizer for Quantum Circuits) \cite{2019voqc} includes five rules for optimizations -- (0) not propagation, (1) Hadamard reduction, (2) single-qubit gate cancellation, (3) two-qubit gate cancellation, and (4) rotation merging. The voqc optimizer applies these rules in the following order in order to optimize a circuit: 0, 1, 3, 2, 3, 1, 2, 4, 3, 2. 

Applying a limited number of optimization rules limits the opportunity to find other optimizations. Instead of defining a set of optimization rules, our compiler is the first to automatically generate circuit identities. Quanto finds an optimized logical quantum circuit that is equivalent to the input logical quantum circuit. 

The first phase in the compiler is generating \textit{circuit identities}. Circuit identities are quantum circuits that are mathematically equivalent. A quantum circuit can be represented as a matrix and two quantum circuits are equivalent when their matrices are equal (for a certain floating point precision). 

Quanto's circuit identity generator enumerates all possible quantum circuits over a given set of gates up to a fixed depth and number of qubits. In the example in Figure \ref{fig:example}, the circuit identity generator would find all possible circuits with two qubits and a depth of five containing the gates \{X, H, CZ\} (or another gate set specified). To efficiently find all such circuit identities, Quanto constructs a hash table where quantum circuits are stored based on the hash of their \textit{unitary matrix}.
The operation of a quantum gate can be represented using a unitary complex matrix. The operation of a quantum circuit can be represented by a unitary matrix obtained from the matrices of individual gates using matrix multiplication and tensor products. In the example in Figure \ref{fig:example}, the unitary matrix is $U = (I \otimes H) \cdot CX \cdot (Z \otimes Z) \cdot CX \cdot (I \otimes H)$.

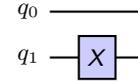
\begin{figure}
    \adjustbox{scale=1}{%
     \begin{tikzcd}
        \lstick{$q_0$} & \qw & \qw \\
        \lstick{$q_1$} & \circuitX & \qw \\
    \end{tikzcd}
    }
\caption{The optimal circuit of the quantum circuit in Figure \ref{fig:example} for the given set}
\label{fig:optimal_example}
\end{figure}

The second phase is the optimizer. When given an input quantum circuit, our optimizer searches for quantum circuits equivalent to the input quantum circuit that minimise the cost function. In this paper we use circuit depth as the cost; that is, a circuit will be replaced by the lowest-depth equivalent circuit known.  In general, however, the cost function can be chosen to optimize any quantifiable criteria of a circuit. For example, a cost function could include the total number of gates, total number of specific gates, such as $T$ gates or two-qubit gates, and/or gate errors. Because we generate all possible circuits, Quanto returns the optimal quantum circuit with respect to the cost function. In the example in Figure \ref{fig:example}, our optimizer returns an equivalent quantum circuit shown in Figure \ref{fig:optimal_example}. Quanto has reduced the depth of the circuit from five to one. This circuit has the minimum depth possible for that given gate set.

\begin{figure}
i) \adjustbox{scale=1}{%
\begin{tikzcd}
    \lstick{$q_0$} & \qw\gategroup[2,steps=3,style={dashed,
    rounded corners,fill=blue!20, inner sep=4pt},
    background]{{\sc}} & \ctrl{1} & \circuitZ & \ctrl{1} & \qw & \qw \\
    \lstick{$q_1$} & \circuitH & \targ{} & \circuitZ & \targ{} & \circuitH & \qw 
\end{tikzcd}
} 

ii) \adjustbox{scale=1}{%
\begin{tikzcd}
    \lstick{$q_0$} & \qw & \ctrl{1}\gategroup[2,steps=3,style={dashed,
    rounded corners,fill=blue!20, inner sep=4pt},
    background]{{\sc}} & \circuitZ & \ctrl{1} & \qw & \qw \\
    \lstick{$q_1$} & \circuitH & \targ{} & \circuitZ & \targ{} & \circuitH & \qw 
\end{tikzcd}
} 

iii) \adjustbox{scale=1}{%
\begin{tikzcd}
    \lstick{$q_0$} & \qw & \ctrl{1} & \circuitZ\gategroup[2,steps=3,style={dashed,
    rounded corners,fill=blue!20, inner sep=4pt},
    background]{{\sc}} & \ctrl{1} & \qw & \qw \\
    \lstick{$q_1$} & \circuitH & \targ{} & \circuitZ & \targ{} & \circuitH & \qw 
\end{tikzcd}
} 
\caption{Splitting the circuit into tiles of width 3 and height 2}
\label{fig:tiles_example}
\end{figure}
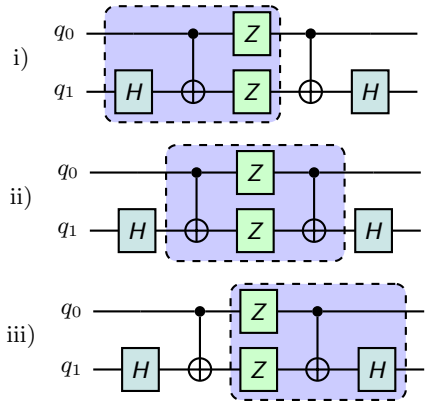

The circuit identities are obtained by simulating the quantum circuits. One problem, however, is that simulating a quantum circuit is expensive if the quantum circuit has a large number of qubits and circuit depth. The promise of quantum computing is dependent on the fact that simulating quantum circuits is classically inefficient. Therefore, the larger the quantum circuit becomes (i.e. more qubits and greater depth), and the more gates that are added, the more configurations there are to generate, which eventually can become infeasible. In order to solve this issue in our optimizer, we introduce \textit{tiles}. A \textit{tile} is a subset of a quantum circuit with a specified length and width. A tile is the source quantum sub-circuit that is matched to equivalent quantum circuits. In the example in Figure \ref{fig:example}, if our tile width is equal to three and our tile length is equal to two, then the tiles in Figure \ref{fig:tiles_example} would be generated.

\begin{figure}
Step 1) \adjustbox{scale=0.65}{%
\begin{tikzcd}
    \lstick{$q_0$} & \qw & \ctrl{1}\gategroup[2,steps=3,style={dashed,
    rounded corners,fill=blue!20, inner sep=4pt},
    background]{{\sc}} & \circuitZ & \ctrl{1} & \qw & \qw \\
    \lstick{$q_1$} & \circuitH & \targ{} & \circuitZ & \targ{} & \circuitH & \qw 
\end{tikzcd}
} $\longrightarrow$
 \adjustbox{scale=0.65}{%
\begin{tikzcd}
    \lstick{$q_0$} & \qw & \qw\gategroup[2,steps=1,style={dashed,
    rounded corners,fill=orange!20, inner sep=4pt},
    background]{{\sc}} & \qw & \qw \\
    \lstick{$q_1$} & \circuitH & \circuitZ  & \circuitH & \qw 
\end{tikzcd}
}

Step 2) \adjustbox{scale=1}{%
\begin{tikzcd}
    \lstick{$q_0$} & \qw\gategroup[2,steps=3,style={dashed,
    rounded corners,fill=blue!20, inner sep=4pt},
    background]{{\sc}} & \qw & \qw & \qw\\
    \lstick{$q_1$} & \circuitH & \circuitZ  & \circuitH & \qw 
\end{tikzcd}
} $\longrightarrow$
\adjustbox{scale=1}{%
\begin{tikzcd}
    \lstick{$q_0$} & \circuitX\gategroup[1,steps=1,style={dashed,
    rounded corners,fill=orange!20, inner sep=4pt},
    background]{{\sc}} & \qw
\end{tikzcd}
}
\caption{Steps to optimize the circuit when the circuit is split into tiles. The boxes shaded in blue are the tiles. The boxes shaded in orange are the substitutions.}
\label{fig:steps_example}
\end{figure}
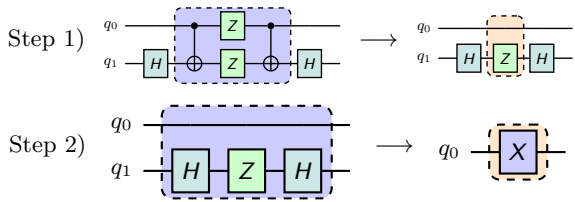

Quanto searches for quantum circuits equivalent to the tiles in the circuit identities database. A cost-based search algorithm is used to apply the substitution, which is a quantum circuit in the circuit identities database that replaces a tile in the quantum circuit. A substitution is the target quantum circuit that defines an equivalent new quantum circuit to replace the matched quantum circuit (i.e. the tile). Figure \ref{fig:steps_example} show how the circuit gets optimized when it is split into tiles.

Our main contribution is presenting Quanto, which is the first quantum optimizer to automatically generate quantum circuit identities. Our automatic generator finds identities used by existing compilers, and in addition finds novel identities (to the best of our knowledge). Quanto can generate novel identities for any given gate set, which is especially useful for quantum hardware with native gate sets that are not standard (for example, the two-qubit iSwap gate is less common than the two-qubit CX gate). In our evaluation we demonstrate that the IBM Qiskit compiler \cite{2017qiskit} can not find these novel identities. 
Quanto also includes a tile-based method to optimize a large quantum circuit using the generated identities. We demonstrate that Quanto's optimizer finds equivalent circuits with smaller depth compared to IBM's Qiskit and Cambridge Quantum Computing's Tket \cite{2020tket} compilers. 

\section{Results}
\subsection{Generating Existing Circuit Identities}
Quanto can automatically discover identities that are known from previous work. \cite{2011equivalentquantumcircuits, 2021gates} We ran the compilers (Quanto, IBM's Qiskit, and Cambridge Quantum Computing's Tket) on a set of quantum circuits with known optimizations. The results show that Quanto is able to find the known optimizations for all of the input quantum circuits, whereas other compilers fail to find some of the identities. The other compilers could find simple identities, such as a circuit with the same gate in a row, but they failed to find more complicated identities. 

These results are shown in Tables \ref{tab:results}, \ref{tab:results2}, \ref{tab:results3}, \ref{tab:results5}, \ref{tab:novel-gates}. The second column of the Tables show the input (uncompiled) circuit and the other columns show the compiled circuit after running on the compiler specified in that column. The last columns in Table \ref{tab:results} and Table \ref{tab:results5} show the circuits that are generated when the uncompiled circuits are run through the Quanto compiler and then the IBM compiler. The reason we show this column is because our optimizer is at the logical level and does not map the qubits to the hardware, a task we leave to the IBM compiler. We show that for the circuits in Table \ref{tab:results} and some of the circuits in Table \ref{tab:results5} the circuits are unchanged by the mapping to the hardware. In particular, it is interesting to look at Circuit A and Circuit C in Table \ref{tab:results5}. One might say that the IBM compiler circuit increased the depth of the uncompiled circuit from depth three to six in Circuit A and from depth three to seven in Circuit C because it needs to map to the hardware, but when we run our compiled circuit (which reduced the depth of the circuit to two in both circuits), the IBM compiler left the optimization unchanged. These results demonstrate that optimizing at the logical level is a useful task that can be done before mapping to the specific quantum hardware.

The Quanto compiled circuits of Circuits B and D in Table \ref{tab:results5} are changed when they are mapped to the hardware. In particular, two-qubit gates that are not applied to neighbouring qubits are changed. In order to address this, we implemented a two-qubit gate neighbouring constraint in Quanto that can be turned on or off. If this constraint is on, it only returns optimized circuits that include two-qubit gates on neighbouring qubits. This is a simple hardware constraint. Quanto could be improved by implementing more specific hardware constraints. The last column in Table \ref{tab:results5} shows the compiled circuits when the constraint is on.

\begin{figure*}
\begin{center}
{\renewcommand{\arraystretch}{1.5}%
\begin{longtable}{ |c|c|c|c|c|c| }
\hline
 \makecell{\thead{Circuit}} & \makecell{\thead{Uncompiled \\ circuit}} & \makecell{\thead{IBM compiled \\ circuit}} & \makecell{\thead{Tket compiled \\ circuit}} & \makecell{\thead{Quanto compiled \\ circuit}} & \makecell{\thead{Quanto-IBM \\ compiled circuit}}\\ 
\specialrule{.3em}{.2em}{.2em}
& & & & &
\\ A
    & 
    \adjustbox{scale=0.60}{%
     \begin{tikzcd}
        \lstick{$q_0$} & \circuitH & \circuitH & \qw \\
    \end{tikzcd}
    }
    
    & 
    \adjustbox{scale=0.60}{%
     \begin{tikzcd}
        \lstick{$q_0$} & \qw & \qw \\
    \end{tikzcd}
    }
    
    & 
    
    \adjustbox{scale=0.60}{%
     \begin{tikzcd}
        \lstick{$q_0$} & \qw & \qw \\
    \end{tikzcd}
    }
    
    & 
    \adjustbox{scale=0.60}{%
     \begin{tikzcd}
        \lstick{$q_0$} & \qw & \qw \\
    \end{tikzcd}
    }
    
    & 
    \adjustbox{scale=0.60}{%
     \begin{tikzcd}
        \lstick{$q_0$} & \qw & \qw \\
    \end{tikzcd}
    }
    \\
& & & & &
\\
\hline
 Depth & 2 & 0 & 0 & 0 & 0 \\
\specialrule{.3em}{.2em}{.2em}
& & & & &
\\ B
    & 
    \adjustbox{scale=0.60}{%
     \begin{tikzcd}
        \lstick{$q_0$} & \circuitT & \gate{T^\dagger} & \qw \\
    \end{tikzcd}
    }
    
    & 
    \adjustbox{scale=0.60}{%
     \begin{tikzcd}
        \lstick{$q_0$} & \qw & \qw \\
    \end{tikzcd}
    }
    
    & 
    
    \adjustbox{scale=0.60}{%
     \begin{tikzcd}
        \lstick{$q_0$} & \qw & \qw \\
    \end{tikzcd}
    }
    
    & 
    \adjustbox{scale=0.60}{%
     \begin{tikzcd}
        \lstick{$q_0$} & \qw & \qw \\
    \end{tikzcd}
    }
    
    & 
    \adjustbox{scale=0.60}{%
     \begin{tikzcd}
        \lstick{$q_0$} & \qw & \qw \\
    \end{tikzcd}
    }
    \\
& & & & &
\\
\hline 
 Depth & 2 & 0 & 0 & 0 & 0 \\
\specialrule{.3em}{.2em}{.2em}
& & & & &
\\ C
    & 
    \adjustbox{scale=0.60}{%
     \begin{tikzcd}
        \lstick{$q_0$} & \circuitH & \ctrl{1} & \circuitH & \qw \\
        \lstick{$q_1$} & \circuitH & \targ{} & \circuitH & \qw \\
    \end{tikzcd}
    }
    
    & 
    \adjustbox{scale=0.60}{%
     \begin{tikzcd}
        \lstick{$q_0$} & \gate{U_2(0,\pi)} & \ctrl{1} & \gate{U_2(0,\pi)} & \qw \\
        \lstick{$q_1$} & \gate{U_2(0,\pi)} & \targ{} & \gate{U_2(0,\pi)} & \qw \\
    \end{tikzcd}
    }
    
    & 
    
    \adjustbox{scale=0.60}{%
     \begin{tikzcd}
        \lstick{$q_0$} & \gate{U_3(\pi/2,0,\pi)} & \ctrl{1} & \gate{U_3(\pi/2,0,\pi)} & \qw \\
        \lstick{$q_1$} & \gate{U_3(\pi/2,0,\pi)} & \targ{} & \gate{U_3(\pi/2,0,\pi)} & \qw \\
    \end{tikzcd}
    }
    
    & 
    \adjustbox{scale=0.60}{%
     \begin{tikzcd}
        \lstick{$q_0$} & \targ{}  & \qw \\
        \lstick{$q_1$} & \ctrl{-1} & \qw \\
    \end{tikzcd}    
    }
    
    & 
    \adjustbox{scale=0.60}{%
     \begin{tikzcd}
        \lstick{$q_0$} & \targ{}  & \qw \\
        \lstick{$q_1$} & \ctrl{-1} & \qw \\
    \end{tikzcd}    
    }
    \\
\hline 
 Depth & 3 & 3 & 3 & 1 & 1 \\
\specialrule{.3em}{.2em}{.2em}
& & & & &
\\
D
   & 
    
    \adjustbox{scale=0.60}{%
     \begin{tikzcd}
        \lstick{$q_0$} & \circuitH & \gate{Z} & \ctrl{1} & \circuitH & \qw \\
        \lstick{$q_1$} & \circuitH & \qw & \targ{} & \circuitH & \qw \\
    \end{tikzcd}
    }
    
    & 
    \adjustbox{scale=0.60}{%
     \begin{tikzcd}
        \lstick{$q_0$} & \gate{U_2(\pi,\pi)} & \ctrl{1} & \gate{U_2(0,\pi)} & \qw \\
        \lstick{$q_1$} & \gate{U_2(0,\pi)} & \targ{} & \gate{U_2(0,\pi)} & \qw \\
    \end{tikzcd}
    }
    
    & 
    
    \adjustbox{scale=0.60}{%
     \begin{tikzcd}
        \lstick{$q_0$} & \gate{U_3(3\pi/2,0,0)} & \ctrl{1} & \gate{U_3(\pi/2,0,\pi)} & \qw \\
        \lstick{$q_1$} & \gate{U_3(\pi/2,0,\pi)} & \targ{} & \gate{U_3(\pi/2,0,\pi)} & \qw \\
    \end{tikzcd}
    }
    
    & 
    
    \adjustbox{scale=0.60}{%
     \begin{tikzcd}
        \lstick{$q_0$} & \circuitX & \targ{} & \qw \\
        \lstick{$q_1$} & \qw & \ctrl{-1} & \qw \\
    \end{tikzcd}    
    }
    
    & 
    
    \adjustbox{scale=0.60}{%
     \begin{tikzcd}
        \lstick{$q_0$} & \gate{U_3(\pi,0,\pi)} & \targ{} & \qw \\
        \lstick{$q_1$} & \qw & \ctrl{-1} & \qw \\
    \end{tikzcd}    
    }
    
    \\
\hline 
 Depth & 4 & 3 & 3 & 2 & 2\\
\specialrule{.3em}{.2em}{.2em}
\caption{\label{tab:results} Comparing Quanto to other compilers for optimizing quantum circuits with known optimizations. IBM and Tket backend is ibmq essex.}
\end{longtable}}
\end{center}
\end{figure*}

\begin{figure*}
\begin{center}
{\renewcommand{\arraystretch}{1.5}%
\begin{longtable}{ |c|c|c|c|c| }
\hline
 \makecell{\thead{Circuit}} & \makecell{\thead{Uncompiled \\ circuit}} & \makecell{\thead{IBM compiled \\ circuit}} & \makecell{\thead{Tket compiled \\ circuit}} & \makecell{\thead{Quanto compiled \\ circuit}}\\ 
\specialrule{.3em}{.2em}{.2em}
& & & &  
\\ 
A
   & 
    \adjustbox{scale=0.60}{%
     \begin{tikzcd}
        \lstick{$q_0$} & \ctrl{1} & \circuitX & \ctrl{1} & \qw \\
        \lstick{$q_1$} & \targ{} & \qw & \targ{} &  \qw \\
    \end{tikzcd}
    }
    
    & 
    \adjustbox{scale=0.60}{%
     \begin{tikzcd}
        \lstick{$q_0$} & \ctrl{1} & \gate{U3(\pi,0,\pi)} & \ctrl{1} & \qw \\
        \lstick{$q_1$} & \targ{} & \qw & \targ{} &  \qw \\
    \end{tikzcd}
    }
    
    & 
    
    \adjustbox{scale=0.60}{%
     \begin{tikzcd}
        \lstick{$q_0$} & \ctrl{1} & \gate{U3(\pi,0,\pi)} & \ctrl{1} & \qw \\
        \lstick{$q_1$} & \targ{} & \qw & \targ{} &  \qw \\
    \end{tikzcd}
    }
    
    & 
    \adjustbox{scale=0.60}{%
     \begin{tikzcd}
        \lstick{$q_0$} & \circuitX & \qw \\
        \lstick{$q_1$} & \circuitX & \qw \\
    \end{tikzcd}
    }
    
    \\
\hline 
 Depth & 3 & 3 & 3 & 1 \\
\specialrule{.3em}{.2em}{.2em}
& & & & 
\\
B
   & 
    
    \adjustbox{scale=0.60}{%
     \begin{tikzcd}
        \lstick{$q_0$} & \ctrl{1} & \qw & \ctrl{1} & \qw \\
        \lstick{$q_1$} & \targ{} & \circuitX & \targ{} &  \qw \\
    \end{tikzcd}
    }
    
    & 
    \adjustbox{scale=0.60}{%
     \begin{tikzcd}
        \lstick{$q_0$} & \ctrl{1} & \qw & \ctrl{1} & \qw \\
        \lstick{$q_1$} & \targ{} & \gate{U3(\pi,0,\pi)} & \targ{} &  \qw \\
    \end{tikzcd}
    }
    
    & 
    \adjustbox{scale=0.60}{%
     \begin{tikzcd}
        \lstick{$q_0$} & \qw & \qw \\
        \lstick{$q_1$} & \gate{U3(\pi,0,\pi)} & \qw \\
    \end{tikzcd}
    }
    
    & 
    \adjustbox{scale=0.60}{%
     \begin{tikzcd}
        \lstick{$q_0$} & \qw & \qw \\
        \lstick{$q_1$} & \circuitX & \qw \\
    \end{tikzcd}
    }
    \\
\hline 
 Depth & 3 & 3 & 1 & 1 \\
\specialrule{.3em}{.2em}{.2em}

& & & & 
\\
C
   & 
    
    \adjustbox{scale=0.60}{%
     \begin{tikzcd}
        \lstick{$q_0$} & \ctrl{1} & \circuitZ & \ctrl{1} & \qw \\
        \lstick{$q_1$} & \targ{} & \qw & \targ{} &  \qw \\
    \end{tikzcd}
    }
    
    & 
    \adjustbox{scale=0.60}{%
     \begin{tikzcd}
        \lstick{$q_0$} & \ctrl{1} & \gate{U1(\pi)} & \ctrl{1} & \qw \\
        \lstick{$q_1$} & \targ{} & \qw & \targ{} &  \qw \\
    \end{tikzcd}
    }
    
    & 
    \adjustbox{scale=0.60}{%
     \begin{tikzcd}
        \lstick{$q_0$} & \gate{U1(\pi)} & \qw \\
        \lstick{$q_1$} & \qw & \qw \\
    \end{tikzcd}
    }
    
    & 
    \adjustbox{scale=0.60}{%
     \begin{tikzcd}
        \lstick{$q_0$} & \circuitZ & \qw \\
        \lstick{$q_1$} & \qw & \qw \\
    \end{tikzcd}
    }

    \\
\hline 
 Depth & 3 & 3 & 1 & 1 \\
\specialrule{.3em}{.2em}{.2em}

& & & & 
\\
D
   & 
    
    \adjustbox{scale=0.60}{%
     \begin{tikzcd}
        \lstick{$q_0$} & \ctrl{1} & \qw & \ctrl{1} & \qw \\
        \lstick{$q_1$} & \targ{} & \circuitZ & \targ{} &  \qw \\
    \end{tikzcd}
    }
    
    & 
    \adjustbox{scale=0.60}{%
     \begin{tikzcd}
        \lstick{$q_0$} & \ctrl{1} & \qw & \ctrl{1} & \qw \\
        \lstick{$q_1$} & \targ{} & \gate{U1(\pi)}  & \targ{} &  \qw \\
    \end{tikzcd}
    }
    
    & 
    
    \adjustbox{scale=0.60}{%
     \begin{tikzcd}
        \lstick{$q_0$} & \ctrl{1} & \qw & \ctrl{1} & \qw \\
        \lstick{$q_1$} & \targ{} & \gate{U1(\pi)}  & \targ{} &  \qw \\
    \end{tikzcd}
    }
    
    & 
    \adjustbox{scale=0.60}{%
     \begin{tikzcd}
        \lstick{$q_0$} & \circuitZ & \qw \\
        \lstick{$q_1$} & \circuitZ & \qw \\
    \end{tikzcd}
    }
    
    \\
\hline 
 Depth & 3 & 3 & 3 & 1 \\
\specialrule{.3em}{.2em}{.2em}

& & & & 
\\
E
   & 
    
    \adjustbox{scale=0.60}{%
     \begin{tikzcd}
        \lstick{$q_0$} & \ctrl{1} & \circuitX & \ctrl{1} & \qw \\
        \lstick{$q_1$} & \control{} & \qw & \control{} &  \qw \\
    \end{tikzcd}
    }
    
    & 
    \adjustbox{scale=0.40}{%
     \begin{tikzcd}
        \lstick{$q_0$} & \qw & \ctrl{1} & \gate{U3(\pi,0,\pi)} & \ctrl{1} \qw & \qw & \qw \\
        \lstick{$q_1$} & \gate{U2(0,\pi)} & \targ{} & \qw & \targ{} & \gate{U2(0,\pi)} & \qw
    \end{tikzcd}
    }
    
    & 
    
    \adjustbox{scale=0.40}{%
     \begin{tikzcd}
        \lstick{$q_0$} & \qw & \ctrl{1} & \gate{U3(\pi,0,\pi)} & \ctrl{1} \qw & \qw & \qw \\
        \lstick{$q_1$} & \gate{U3(\pi/2,0,\pi)} & \targ{} & \qw & \targ{} & \gate{U3(\pi/2,0,\pi)} & \qw
    \end{tikzcd}
    }
    
    & 
    \adjustbox{scale=0.60}{%
     \begin{tikzcd}
        \lstick{$q_0$} & \circuitX & \qw \\
        \lstick{$q_1$} & \circuitZ & \qw \\
    \end{tikzcd}
    }

    \\
\hline 
 Depth & 3 & 5 & 5 & 1 \\
\specialrule{.3em}{.2em}{.2em}

& & & & 
\\
F
   & 
    
    \adjustbox{scale=0.60}{%
     \begin{tikzcd}
        \lstick{$q_0$} & \ctrl{1} & \qw & \ctrl{1} & \qw \\
        \lstick{$q_1$} & \control{} & \circuitX & \control{} &  \qw \\
    \end{tikzcd}
    }
    
    & 
    \adjustbox{scale=0.40}{%
     \begin{tikzcd}
        \lstick{$q_0$} & \qw & \ctrl{1} & \qw & \ctrl{1} \qw & \qw & \qw \\
        \lstick{$q_1$} & \gate{U2(0,\pi)} & \targ{} & \gate{U1(3\pi)} & \targ{} & \gate{U2(0,\pi)} & \qw
    \end{tikzcd}
    }
    
    & 
    
    \adjustbox{scale=0.40}{%
     \begin{tikzcd}
        \lstick{$q_0$} & \qw & \ctrl{1} & \qw & \ctrl{1} \qw & \qw & \qw \\
        \lstick{$q_1$} & \gate{U3(\pi/2,0,\pi)} & \targ{} & \gate{U1(\pi)}  & \targ{} &  \gate{U3(\pi/2,0,\pi)} & \qw
    \end{tikzcd}
    }
    
    & 
    \adjustbox{scale=0.60}{%
     \begin{tikzcd}
        \lstick{$q_0$} & \circuitZ & \qw \\
        \lstick{$q_1$} & \circuitX & \qw \\
    \end{tikzcd}
    }
    \\
\hline 
 Depth & 3 & 5 & 5 & 1 \\
\specialrule{.3em}{.2em}{.2em}
\caption{\label{tab:results2} Comparing Quanto to other compilers for optimizing quantum circuits with known optimizations. IBM and Tket backend is Aerbackend.}
\end{longtable}}
\end{center}
\end{figure*}

\begin{figure*}
\begin{center}
{\renewcommand{\arraystretch}{1.5}%
\begin{longtable}{ |c|c|c|c|c| }
\hline
 \makecell{\thead{Circuit}} & \makecell{\thead{Uncompiled \\ circuit}} & \makecell{\thead{IBM compiled \\ circuit}} & \makecell{\thead{Tket compiled \\ circuit}} & \makecell{\thead{Quanto compiled \\ circuit}}\\ 
\specialrule{.3em}{.2em}{.2em}
& & & & 
\\ 
A
   & 
    \adjustbox{scale=0.60}{%
     \begin{tikzcd}
        \lstick{$q_0$} & \ctrl{1} & \circuitZ & \ctrl{1} & \qw \\
        \lstick{$q_1$} & \control{} & \qw & \control{} &  \qw \\
    \end{tikzcd}
    }
    
    & 
    \adjustbox{scale=0.40}{%
     \begin{tikzcd}
        \lstick{$q_0$} & \qw & \ctrl{1} & \gate{U1(\pi)} & \ctrl{1} \qw & \qw & \qw \\
        \lstick{$q_1$} & \gate{U2(0,\pi)} & \targ{} & \qw & \targ{} & \gate{U2(0,\pi)} & \qw
    \end{tikzcd}
    }
    
    & 
    \adjustbox{scale=0.60}{%
     \begin{tikzcd}
        \lstick{$q_0$} & \gate{U1(\pi)} & \qw \\
        \lstick{$q_1$} & \qw & \qw \\
    \end{tikzcd}
    }
    
    & 
    \adjustbox{scale=0.60}{%
     \begin{tikzcd}
        \lstick{$q_0$} & \circuitZ & \qw \\
        \lstick{$q_1$} & \qw & \qw \\
    \end{tikzcd}
    }
    
    \\
\hline 
 Depth & 3 & 3 & 1 & 1 \\
\specialrule{.3em}{.2em}{.2em}
& & & & 
\\ 
B
   & 
    \adjustbox{scale=0.60}{%
     \begin{tikzcd}
        \lstick{$q_0$} & \ctrl{1} & \qw & \ctrl{1} & \qw \\
        \lstick{$q_1$} & \control{} & \circuitZ & \control{} &  \qw \\
    \end{tikzcd}
    }
    
    & 
    \adjustbox{scale=0.40}{%
     \begin{tikzcd}
        \lstick{$q_0$} & \qw & \ctrl{1} & \qw & \ctrl{1} \qw & \qw & \qw \\
        \lstick{$q_1$} & \gate{U2(0,\pi)} & \targ{} & \gate{U3(-\pi,\pi/2,3\pi/2)} & \targ{} & \gate{U2(0,\pi)} & \qw
    \end{tikzcd}
    }
    
    & 
    
    \adjustbox{scale=0.40}{%
     \begin{tikzcd}
        \lstick{$q_0$} & \qw & \ctrl{1} & \qw & \ctrl{1} \qw & \qw & \qw \\
        \lstick{$q_1$} & \gate{U3(\pi/2,0,\pi)} & \targ{} & \gate{U3(3\pi,0,\pi)} & \targ{} & \gate{U3(\pi/2,0,\pi)} & \qw
    \end{tikzcd}
    }
    
    & 
    \adjustbox{scale=0.60}{%
     \begin{tikzcd}
        \lstick{$q_0$} & \qw & \qw \\
        \lstick{$q_1$} & \circuitZ & \qw \\
    \end{tikzcd}
    }

    \\
\hline 
 Depth & 3 & 5 & 5 & 1 \\
\specialrule{.3em}{.2em}{.2em}

& & & & 
\\ 
C
   & 
    \adjustbox{scale=0.40}{%
     \begin{tikzcd}
        \lstick{$q_0$} & \qw & \ctrl{1} & \qw & \ctrl{1} & \qw & \qw \\
        \lstick{$q_1$} & \circuitH & \targ{} & \circuitX & \targ{} & \circuitH & \qw \\
    \end{tikzcd}
    }
    
    & 
    \adjustbox{scale=0.40}{%
     \begin{tikzcd}
        \lstick{$q_0$} & \qw & \ctrl{1} & \qw & \ctrl{1} \qw & \qw & \qw \\
        \lstick{$q_1$} & \gate{U2(0,\pi)} & \targ{} & \gate{U3(\pi,0,\pi)} & \targ{} & \gate{U2(0,\pi)} & \qw \\
    \end{tikzcd}
    }
    
    & 
    
    \adjustbox{scale=0.60}{%
     \begin{tikzcd}
        \lstick{$q_0$} & \qw & \qw \\
        \lstick{$q_1$} & \gate{U1(\pi)} & \qw \\
    \end{tikzcd}
    }
    
    & 
    \adjustbox{scale=0.60}{%
     \begin{tikzcd}
        \lstick{$q_0$} & \qw & \qw \\
        \lstick{$q_1$} & \circuitZ & \qw \\
    \end{tikzcd}
    }
    
    \\
\hline 
 Depth & 5 & 5 & 1 & 1 \\
\specialrule{.3em}{.2em}{.2em}

& & & & 
\\ 
D
   & 
    \adjustbox{scale=0.40}{%
     \begin{tikzcd}
        \lstick{$q_0$} & \qw & \ctrl{1} & \circuitZ & \ctrl{1} & \qw & \qw \\
        \lstick{$q_1$} & \circuitH & \targ{} & \circuitZ & \targ{} & \circuitH & \qw \\
    \end{tikzcd}
    }
    
    & 
    \adjustbox{scale=0.40}{%
     \begin{tikzcd}
        \lstick{$q_0$} & \qw & \ctrl{1} & \gate{U1(\pi)} & \ctrl{1} & \qw & \qw \\
        \lstick{$q_1$} & \gate{U2(0,\pi)} & \targ{} & \gate{U1(\pi)} & \targ{} & \gate{U2(0,\pi)} & \qw \\
    \end{tikzcd}
    }
    
    & 
    
    \adjustbox{scale=0.40}{%
     \begin{tikzcd}
        \lstick{$q_0$} & \gate{U1(\pi)} & \ctrl{1} & \qw & \ctrl{1} \qw & \qw & \qw \\
        \lstick{$q_1$} & \gate{U3(\pi/2,0,\pi)} & \targ{} & \gate{U1(\pi)} & \targ{} & \gate{U3(\pi/2,0,\pi)} & \qw \\
    \end{tikzcd}
    }
    
    & 
    \adjustbox{scale=0.60}{%
     \begin{tikzcd}
        \lstick{$q_0$} & \qw & \qw \\
        \lstick{$q_1$} & \circuitX & \qw \\
    \end{tikzcd}
    }
    
    \\
\hline 
 Depth & 5 & 5 & 5 & 1 \\
\specialrule{.3em}{.2em}{.2em}
\caption{\label{tab:results3} Comparing Quanto to other compilers for optimizing quantum circuits with known optimizations. IBM and Tket backend is Aerbackend.}
\end{longtable}}
\end{center}
\end{figure*}

\begin{figure*}
\begin{center}
{\renewcommand{\arraystretch}{1.5}%
\begin{longtable}{ |c|c|c|c|c|c|c| }
\hline
 \makecell{\thead{Circuit}} & \makecell{\thead{Uncompiled \\ circuit}} & \makecell{\thead{IBM compiled \\ circuit}} & \makecell{\thead{Tket compiled \\ circuit}} & \makecell{\thead{Quanto compiled \\ circuit}} & \makecell{\thead{Quanto-IBM \\ compiled circuit}} & \makecell{\thead{Quanto compiled \\  circuit \\ (with constraint)}}\\ 
\specialrule{.3em}{.2em}{.2em}
& & & & & &
\\
A
    & 
    
    \adjustbox{scale=0.45}{%
     \begin{tikzcd}
        \lstick{$q_0$} & \qw & \ctrl{1} & \ctrl{2} & \qw \\
        \lstick{$q_1$} & \ctrl{1} & \targ{} & \qw & \qw \\
        \lstick{$q_2$} & \targ{} & \qw & \targ{} & \qw \\
    \end{tikzcd}
    }
    
    & 
    \adjustbox{scale=0.45}{%
     \begin{tikzcd}
        \lstick{$q_0$} & \targ{} & \qw & \ctrl{1} & \targ{} & \ctrl{1} & \qw & \qw \\
        \lstick{$q_1$} & \ctrl{-1} & \targ{} & \targ{} & \ctrl{-1} & \targ{} & \targ{} & \qw \\
        \lstick{$q_2$} & \qw & \ctrl{-1} & \qw & \qw & \qw & \ctrl{-1} & \qw \\
    \end{tikzcd}
    }
    
    & 
    
    \adjustbox{scale=0.45}{%
     \begin{tikzcd}
        \lstick{$q_0$} & \qw & \ctrl{1} & \ctrl{2} & \qw \\
        \lstick{$q_1$} & \ctrl{1} & \targ{} & \qw & \qw \\
        \lstick{$q_2$} & \targ{} & \qw & \targ{} & \qw \\
    \end{tikzcd}
    }
    
    & 
    
    \adjustbox{scale=0.45}{%
     \begin{tikzcd}
        \lstick{$q_0$} & \ctrl{1} & \qw & \qw \\
        \lstick{$q_1$} & \targ{} & \ctrl{1} & \qw \\
        \lstick{$q_2$} & \qw & \targ{} & \qw \\
    \end{tikzcd}    
    }
    
    & 
    
    \adjustbox{scale=0.45}{%
     \begin{tikzcd}
        \lstick{$q_0$} & \ctrl{1} & \qw & \qw \\
        \lstick{$q_1$} & \targ{} & \ctrl{1} & \qw \\
        \lstick{$q_2$} & \qw & \targ{} & \qw \\
    \end{tikzcd}    
    }
    
    & 
    
    \adjustbox{scale=0.45}{%
     \begin{tikzcd}
        \lstick{$q_0$} & \ctrl{1} & \qw & \qw \\
        \lstick{$q_1$} & \targ{} & \ctrl{1} & \qw \\
        \lstick{$q_2$} & \qw & \targ{} & \qw \\
    \end{tikzcd}    
    }

    \\
\hline 
 Depth & 3 & 6 & 3 & 2 & 2 & 2\\
\specialrule{.3em}{.2em}{.2em}
& & & & & &
\\
B
   & 
    \adjustbox{scale=0.45}{%
     \begin{tikzcd}
        \lstick{$q_0$} & \ctrl{1} & \qw & \ctrl{1} & \qw & \qw \\
        \lstick{$q_1$} & \targ{} & \ctrl{1} & \targ{} & \ctrl{1} & \qw \\
        \lstick{$q_2$} & \qw & \targ{} & \qw & \targ{} & \qw \\
    \end{tikzcd}
    }
    
    & 
     \adjustbox{scale=0.45}{%
     \begin{tikzcd}
        \lstick{$q_0$} & \ctrl{1} & \qw & \ctrl{1} & \qw & \qw \\
        \lstick{$q_1$} & \targ{} & \ctrl{1} & \targ{} & \ctrl{1} & \qw \\
        \lstick{$q_2$} & \qw & \targ{} & \qw & \targ{} & \qw \\
    \end{tikzcd}
    }  
    
    & 
    
     \adjustbox{scale=0.45}{%
     \begin{tikzcd}
        \lstick{$q_0$} & \ctrl{1} & \qw & \ctrl{1} & \qw & \qw \\
        \lstick{$q_1$} & \targ{} & \ctrl{1} & \targ{} & \ctrl{1} & \qw \\
        \lstick{$q_2$} & \qw & \targ{} & \qw & \targ{} & \qw \\
    \end{tikzcd}
    }   
    
    & 
    \adjustbox{scale=0.45}{%
     \begin{tikzcd}
        \lstick{$q_0$} & \ctrl{2} & \qw \\
        \lstick{$q_1$} & \qw & \qw  \\
        \lstick{$q_2$} & \targ{} & \qw \\
    \end{tikzcd}
    }   
    
    & 
    \adjustbox{scale=0.45}{%
     \begin{tikzcd}
        \lstick{$q_0$} & \qw      & \qw       & \qw       & \targ{}  & \qw \\
        \lstick{$q_1$} & \ctrl{1} & \targ{}   & \ctrl{1}  & \ctrl{-1}  & \qw\\
        \lstick{$q_2$} & \targ{}  & \ctrl{-1} & \targ{}   & \qw  & \qw \\
    \end{tikzcd}
    } 
    
    & 
    
     \adjustbox{scale=0.45}{%
     \begin{tikzcd}
        \lstick{$q_0$} & \ctrl{1} & \qw & \ctrl{1} & \qw & \qw \\
        \lstick{$q_1$} & \targ{} & \ctrl{1} & \targ{} & \ctrl{1} & \qw \\
        \lstick{$q_2$} & \qw & \targ{} & \qw & \targ{} & \qw \\
    \end{tikzcd}
    }  
 \\
\hline 
 Depth & 4 & 4 & 4 & 1 & 4 & 4\\
\specialrule{.3em}{.2em}{.2em}

& & & & & &
\\
C
   & 
    \adjustbox{scale=0.45}{%
     \begin{tikzcd}
        \lstick{$q_0$} & \ctrl{2} & \qw & \ctrl{1} & \qw \\
        \lstick{$q_1$} & \qw & \ctrl{1} & \targ{} & \qw \\
        \lstick{$q_2$} & \targ{} & \targ{} & \qw & \qw \\
    \end{tikzcd}
    }
    
    & 
    \adjustbox{scale=0.45}{%
     \begin{tikzcd}
        \lstick{$q_0$} & \ctrl{1}  & \targ{}   & \ctrl{1} & \qw      & \targ{}   & \ctrl{1} & \qw      & \qw \\
        \lstick{$q_1$} & \targ{}   & \ctrl{-1} & \targ{}  & \ctrl{1} & \ctrl{-1} & \targ{}  & \ctrl{1} & \qw \\
        \lstick{$q_2$} & \qw       & \qw       & \qw      & \targ{}  & \qw       & \qw      & \targ{}  & \qw \\
    \end{tikzcd}
    }
    
    & 
    
    \adjustbox{scale=0.45}{%
     \begin{tikzcd}
        \lstick{$q_0$} & \ctrl{2} & \qw & \ctrl{1} & \qw \\
        \lstick{$q_1$} & \qw & \ctrl{1} & \targ{} & \qw \\
        \lstick{$q_2$} & \targ{} & \targ{} & \qw & \qw \\
    \end{tikzcd}
    }  
    
    & 
    \adjustbox{scale=0.45}{%
     \begin{tikzcd}
        \lstick{$q_0$} & \ctrl{1} & \qw & \qw \\
        \lstick{$q_1$} & \targ{} & \ctrl{1} & \qw \\
        \lstick{$q_2$} & \qw & \targ{} & \qw \\
    \end{tikzcd}
    }    
    
    & 
    \adjustbox{scale=0.45}{%
     \begin{tikzcd}
        \lstick{$q_0$} & \ctrl{1} & \qw & \qw \\
        \lstick{$q_1$} & \targ{} & \ctrl{1} & \qw \\
        \lstick{$q_2$} & \qw & \targ{} & \qw \\
    \end{tikzcd}
    }   
    
    & 
    \adjustbox{scale=0.45}{%
     \begin{tikzcd}
        \lstick{$q_0$} & \ctrl{1} & \qw & \qw \\
        \lstick{$q_1$} & \targ{} & \ctrl{1} & \qw \\
        \lstick{$q_2$} & \qw & \targ{} & \qw \\
    \end{tikzcd}
    }  
    
    \\
\hline 
 Depth & 3 & 7 & 3 & 2 & 2 & 2\\
\specialrule{.3em}{.2em}{.2em}

& & & & & &
\\
D
   & 
    \adjustbox{scale=0.45}{%
     \begin{tikzcd}
        \lstick{$q_0$} & \qw & \ctrl{1} & \qw & \qw \\
        \lstick{$q_1$} & \ctrl{1} & \targ{} & \ctrl{1} & \qw \\
        \lstick{$q_2$} & \targ{} & \qw & \targ{} & \qw \\
    \end{tikzcd}
    }
    
    & 
    \adjustbox{scale=0.45}{%
     \begin{tikzcd}
        \lstick{$q_0$} & \qw & \ctrl{1} & \qw & \qw \\
        \lstick{$q_1$} & \ctrl{1} & \targ{} & \ctrl{1} & \qw \\
        \lstick{$q_2$} & \targ{} & \qw & \targ{} & \qw \\
    \end{tikzcd}
    }
    
    & 
    
    \adjustbox{scale=0.45}{%
     \begin{tikzcd}
        \lstick{$q_0$} & \qw & \ctrl{1} & \qw & \qw \\
        \lstick{$q_1$} & \ctrl{1} & \targ{} & \ctrl{1} & \qw \\
        \lstick{$q_2$} & \targ{} & \qw & \targ{} & \qw \\
    \end{tikzcd}
    } 
    
    & 
    \adjustbox{scale=0.45}{%
     \begin{tikzcd}
        \lstick{$q_0$} & \ctrl{1} & \ctrl{2} & \qw \\
        \lstick{$q_1$} & \targ{} & \qw & \qw \\
        \lstick{$q_2$} & \qw & \targ{} & \qw \\
    \end{tikzcd}
    }  
    
    & 
    \adjustbox{scale=0.45}{%
     \begin{tikzcd}
        \lstick{$q_0$} & \qw      & \qw       & \qw       & \qw  & \targ{} & \qw \\
        \lstick{$q_1$} & \targ{} & \ctrl{1}   & \targ{}  & \ctrl{1}  & \ctrl{-1} & \qw \\
        \lstick{$q_2$} & \ctrl{-1}  & \targ{} & \ctrl{-1}   & \targ{}  & \qw & \qw \\
    \end{tikzcd}
    }

    & 
    
    \adjustbox{scale=0.45}{%
     \begin{tikzcd}
        \lstick{$q_0$} & \qw & \ctrl{1} & \qw & \qw \\
        \lstick{$q_1$} & \ctrl{1} & \targ{} & \ctrl{1} & \qw \\
        \lstick{$q_2$} & \targ{} & \qw & \targ{} & \qw \\
    \end{tikzcd}
    } 
      \\
\hline 
 Depth & 3 & 3 & 3 & 2 & 5 & 3\\
\specialrule{.3em}{.2em}{.2em}
& & & & & &
\\
E
    & 

    \adjustbox{scale=0.25}{%
     \begin{tikzcd}
        \lstick{$q_0$} & \qw & \ctrl{1} & \ctrl{2} & \qw & \qw & \ctrl{1} & \ctrl{2} & \qw & \qw & \ctrl{1} & \ctrl{2} & \qw\\
        \lstick{$q_1$} & \ctrl{1} & \targ{} & \qw & \qw & \ctrl{1} & \targ{} & \qw & \qw & \ctrl{1} & \targ{} & \qw & \qw\\
        \lstick{$q_2$} & \targ{} & \qw & \targ{} & \qw & \targ{} & \qw & \targ{} & \qw & \targ{} & \qw & \targ{} & \qw\\
        
        \lstick{$q_3$} & \qw & \ctrl{1} & \ctrl{2} & \qw & \qw & \ctrl{1} & \ctrl{2} & \qw & \qw & \ctrl{1} & \ctrl{2} & \qw\\
        \lstick{$q_4$} & \ctrl{1} & \targ{} & \qw & \qw & \ctrl{1} & \targ{} & \qw & \qw & \ctrl{1} & \targ{} & \qw & \qw\\
        \lstick{$q_5$} & \targ{} & \qw & \targ{} & \qw & \targ{} & \qw & \targ{} & \qw & \targ{} & \qw & \targ{} & \qw\\
    \end{tikzcd}
    } 
    
    &
    
    \makecell{Circuit too large \\ for displaying image}

    &

    \adjustbox{scale=0.25}{%
     \begin{tikzcd}
        \lstick{$q_0$} & \qw & \ctrl{1} & \ctrl{2} & \qw & \qw & \ctrl{1} & \ctrl{2} & \qw & \qw & \ctrl{1} & \ctrl{2} & \qw\\
        \lstick{$q_1$} & \ctrl{1} & \targ{} & \qw & \qw & \ctrl{1} & \targ{} & \qw & \qw & \ctrl{1} & \targ{} & \qw & \qw\\
        \lstick{$q_2$} & \targ{} & \qw & \targ{} & \qw & \targ{} & \qw & \targ{} & \qw & \targ{} & \qw & \targ{} & \qw\\
        
        \lstick{$q_3$} & \qw & \ctrl{1} & \ctrl{2} & \qw & \qw & \ctrl{1} & \ctrl{2} & \qw & \qw & \ctrl{1} & \ctrl{2} & \qw\\
        \lstick{$q_4$} & \ctrl{1} & \targ{} & \qw & \qw & \ctrl{1} & \targ{} & \qw & \qw & \ctrl{1} & \targ{} & \qw & \qw\\
        \lstick{$q_5$} & \targ{} & \qw & \targ{} & \qw & \targ{} & \qw & \targ{} & \qw & \targ{} & \qw & \targ{} & \qw\\
    \end{tikzcd}
    } 
    
    & 
    
    \adjustbox{scale=0.45}{%
     \begin{tikzcd}
        \lstick{$q_0$} & \qw      & \ctrl{1} & \qw \\
        \lstick{$q_1$} & \ctrl{1} & \targ{}  & \qw \\
        \lstick{$q_2$} & \targ{}  & \qw{} & \qw \\
        \lstick{$q_3$} & \qw & \ctrl{1}  & \qw \\
        \lstick{$q_4$} & \ctrl{1}  & \targ{}      & \qw \\
        \lstick{$q_5$} & \targ{}  & \qw      & \qw \\
    \end{tikzcd}    
    } 
    
    & 
    
    \adjustbox{scale=0.45}{%
     \begin{tikzcd}
        \lstick{$q_0$} & \qw      & \ctrl{1} & \qw \\
        \lstick{$q_1$} & \ctrl{1} & \targ{}  & \qw \\
        \lstick{$q_2$} & \targ{}  & \qw{} & \qw \\
        \lstick{$q_3$} & \qw & \ctrl{1}  & \qw \\
        \lstick{$q_4$} & \ctrl{1}  & \targ{}      & \qw \\
        \lstick{$q_5$} & \targ{}  & \qw      & \qw \\
    \end{tikzcd}    
    } 
    
    & 
    
    \adjustbox{scale=0.45}{%
     \begin{tikzcd}
        \lstick{$q_0$} & \qw      & \ctrl{1} & \qw \\
        \lstick{$q_1$} & \ctrl{1} & \targ{}  & \qw \\
        \lstick{$q_2$} & \targ{}  & \qw{} & \qw \\
        \lstick{$q_3$} & \qw & \ctrl{1}  & \qw \\
        \lstick{$q_4$} & \ctrl{1}  & \targ{}      & \qw \\
        \lstick{$q_5$} & \targ{}  & \qw      & \qw \\
    \end{tikzcd}    
    } 
    \\
    
\hline 
 Depth & 9 & 31 & 9 & 2 & 2 & 2\\
\specialrule{.3em}{.2em}{.2em}
\caption{\label{tab:results5} Comparing Quanto to other compilers for optimizing quantum circuits with known optimizations. IBM backend is ibmq lima and Tket backend is Aerbackend.}
\end{longtable}}
\end{center}
\end{figure*}

\subsection{Generating Novel Circuit Identities}
Quanto can also generate new identities, especially when given custom gates, as it is more likely that identities for custom gates have not been previously discovered. Existing compilers have a fixed repertoire of identities using common gates. The advantage of Quanto is that it can find identities for any custom gate, leading to optimizations that may be missed by other compilers. In Table \ref{tab:novel-identities} we show some of these novel identities (these identities were found from our identity database and we are not aware of any previous reports of these identities). We compare our novel identity optimizations to the IBM Qiskit compiler, as it allows us to specify basis gates for compiling. IBM's compiler is unable to find these circuit identities. 

\subsection{Optimizing Quantum Circuits}
Other than generating existing and novel identities, Quanto can also optimize any quantum circuit as shown in Table \ref{tab:results}.

\begin{figure*}
\begin{center}

{\renewcommand{\arraystretch}{1.5}%
\begin{longtable}{ |c|c|c|c| }
\hline
 \makecell{\thead{Circuit}} & \makecell{\thead{Uncompiled \\ circuit}} & \makecell{\thead{IBM compiled \\ circuit}} & \makecell{\thead{Quanto compiled \\ circuit}} \\ 
\specialrule{.3em}{.2em}{.2em}
& & & 
\\ A &

    \adjustbox{scale=0.62}{%
     \begin{tikzcd}
        \lstick{$q_0$} & \qw & \ctrl{1} & \qw & \qw \\
        \lstick{$q_1$} & \gate{S^\dagger} & \targ{} & \circuitS & \qw \\
    \end{tikzcd}
    }
    
    &
    \adjustbox{scale=0.62}{%
     \begin{tikzcd}
        \lstick{$q_0$} & \qw & \ctrl{1} & \qw & \qw \\
        \lstick{$q_1$} & \gate{U1(-\pi/2)} & \targ{} & \gate{U1(\pi/2)} & \qw \\
    \end{tikzcd}
    }
    &
    \adjustbox{scale=0.62}{%
     \begin{tikzcd}
        \lstick{$q_0$} & \ctrl{1} & \qw \\
        \lstick{$q_1$} & \gate{Y} & \qw \\
    \end{tikzcd}
    }

    \\
\hline 
 Depth & 3 & 3 & 1 \\
\specialrule{.3em}{.2em}{.2em}
& & & 
\\ B
    & 
        \adjustbox{scale=0.62}{%
     \begin{tikzcd}
        \lstick{$q_0$} & \qw & \ctrl{1} & \qw & \qw \\
        \lstick{$q_1$} & \circuitH & \targ{} & \circuitH & \qw \\
    \end{tikzcd}
    }
    
    &
    \adjustbox{scale=0.62}{%
     \begin{tikzcd}
        \lstick{$q_0$} & \ctrl{1} & \qw \\
        \lstick{$q_1$} & \control{} & \qw \\
    \end{tikzcd}
    }   
    &
    \adjustbox{scale=0.62}{%
     \begin{tikzcd}
        \lstick{$q_0$} & \ctrl{1} & \qw \\
        \lstick{$q_1$} & \control{} & \qw \\
    \end{tikzcd}
    }

    \\
\hline 
 Depth & 3 & 1 & 1 \\
\specialrule{.3em}{.2em}{.2em}

& & & 
\\ C
    & 
    \adjustbox{scale=0.62}{%
     \begin{tikzcd}
        \lstick{$q_0$} & \circuitS & \circuitH & \ctrl{1} & \targ{} & \qw & \qw \\
        \lstick{$q_1$} & \circuitS & \qw & \targ{} & \ctrl{-1} & \circuitH & \qw \\
    \end{tikzcd}
    }
    
    &
    \adjustbox{scale=0.62}{%
     \begin{tikzcd}
        \lstick{$q_0$} & \gate{U2(0,3\pi/2)} & \ctrl{1} & \targ{} & \qw & \qw \\
        \lstick{$q_1$} & \gate{U1(\pi/2)} & \targ{} & \ctrl{-1} & \gate{U2(0,\pi)} & \qw \\
    \end{tikzcd}
    }
    
    &
    \adjustbox{scale=0.62}{%
     \begin{tikzcd}
        \lstick{$q_0$} & \gate[2]{\text{iSWAP}} & \qw \\
        \lstick{$q_1$} & & \qw \\
    \end{tikzcd}
    }

    \\
\hline 
 Depth & 5 & 4 & 1 \\
\specialrule{.3em}{.2em}{.2em}
\caption{\label{tab:novel-gates} Comparing Quanto to other compilers for optimizing quantum circuits for unique gate sets}
\end{longtable}}
\end{center}
\end{figure*}

\begin{figure*}
\begin{center}

{\renewcommand{\arraystretch}{1.5}%
\begin{longtable}{ |c|c|c|c| }
\hline
 \makecell{\thead{Circuit}} & \makecell{\thead{Uncompiled \\ circuit}} & \makecell{\thead{IBM compiled \\ circuit}} & \makecell{\thead{Quanto compiled \\ circuit}} \\ 
\Xhline{4\arrayrulewidth}
& & & 
\\ A &

    \adjustbox{scale=0.62}{%
     \begin{tikzcd}
        \lstick{$q_0$} & \qw & \targ{} & \ctrl{1} & \gate{S^\dagger} & \qw \\
        \lstick{$q_1$} & \circuitS & \ctrl{-1} & \targ{} & \circuitS & \qw \\
    \end{tikzcd}
    }
    
    &
    \adjustbox{scale=0.62}{%
     \begin{tikzcd}
        \lstick{$q_0$} & \qw & \targ{} & \ctrl{1} & \gate{U1(-\pi/2)} & \qw \\
        \lstick{$q_1$} & \gate{U1(\pi/2)} & \ctrl{-1} & \targ{} & \gate{U1(\pi/2)} & \qw \\
    \end{tikzcd}
    }
    &
    \adjustbox{scale=0.62}{%
     \begin{tikzcd}
        \lstick{$q_0$} & \qw & \ctrl{1} & \targ{} & \ctrl{1} & \qw \\
        \lstick{$q_1$} & \qw & \control{} & \ctrl{-1} & \targ{}  & \qw \\
    \end{tikzcd}
    }

    \\
\hline 
 Depth & 4 & 4 & 3 \\
\Xhline{4\arrayrulewidth}
& & & 
\\ B
    & 
    \adjustbox{scale=0.62}{%
     \begin{tikzcd}
        \lstick{$q_0$} & \circuitY & \ctrl{1} & \circuitX & \targ{} & \qw \\
        \lstick{$q_1$} & \qw & \control{} & \circuitY & \ctrl{-1} & \qw \\
    \end{tikzcd}
    }
    
    & 
    \adjustbox{scale=0.62}{%
     \begin{tikzcd}
        \lstick{$q_0$} & \gate{U_3(\pi,\pi/2,\pi/2)} & \ctrl{1} & \gate{U_3(\pi,0,\pi)} & \targ{} & \qw \\
        \lstick{$q_1$} & \qw & \control{} & \gate{U_3(\pi,\pi/2,\pi/2)} & \ctrl{-1} & \qw \\
    \end{tikzcd}
    }
    
    & 
    
    \adjustbox{scale=0.62}{%
     \begin{tikzcd}
        \lstick{$q_0$} & \ctrl{1}   & \targ{}  & \circuitY & \qw \\
        \lstick{$q_1$} & \control{} & \ctrl{-1} & \circuitY & \qw \\
    \end{tikzcd}
    }

    \\
\hline 
 Depth & 4 & 4 & 3  \\
\Xhline{4\arrayrulewidth}

& & & 
\\ C
    & 
    \adjustbox{scale=0.62}{%
     \begin{tikzcd}
        \lstick{$q_0$} & \circuitZ & \gate[2]{\text{iSWAP}} & \qw & \qw \\
        \lstick{$q_1$} & \circuitX & & \circuitH & \qw \\
    \end{tikzcd}
    }   
    
    & 
    \adjustbox{scale=0.62}{%
     \begin{tikzcd}
        \lstick{$q_0$} & \gate{U1(\pi)} & \gate[2]{\text{iSWAP}} & \qw & \qw \\
        \lstick{$q_1$} & \gate{U3(\pi,0,\pi)} & & \gate{U2(0,\pi)} & \qw \\
    \end{tikzcd}
    }  
    
    & 
    
    \adjustbox{scale=0.62}{%
     \begin{tikzcd}
        \lstick{$q_0$} & \gate[2]{\text{iSWAP}} & \circuitY & \qw \\
        \lstick{$q_1$} & & \circuitH & \qw \\
    \end{tikzcd}
    }

    \\
\hline 
 Depth & 3 & 3 & 2  \\
\Xhline{4\arrayrulewidth}

& & & 
\\ D
    & 
    \adjustbox{scale=0.62}{%
     \begin{tikzcd}
        \lstick{$q_0$} & \qw & \ctrl{1} & \gate[2]{\text{iSWAP}} & \qw \\
        \lstick{$q_1$} & \gate{S^\dagger} & \targ{} & & \qw \\
    \end{tikzcd}
    }  = 
    \newline
 \adjustbox{scale=0.62}{%
     \begin{tikzcd}
        \lstick{$q_0$} & \qw & \ctrl{1}             & \circuitS & \circuitH & \ctrl{1} & \targ{} & \qw & \qw \\
        \lstick{$q_1$} & \gate{S^\dagger} & \targ{} & \circuitS & \qw & \targ{} & \ctrl{-1} & \circuitH & \qw \\
    \end{tikzcd}
    }
    
    & 
    \adjustbox{scale=0.62}{%
     \begin{tikzcd}
        \lstick{$q_0$} & \qw & \ctrl{1}             & \gate{U2(0,3\pi/2)} & \ctrl{1} & \targ{} & \qw & \qw \\
        \lstick{$q_1$} & \gate{U1(-\pi/2)} & \targ{} & \gate{U1(\pi/2)} & \targ{} & \ctrl{-1} & \gate{U2(0,\pi)} & \qw & \qw \\
    \end{tikzcd}
    }
    
    & 
    
    \adjustbox{scale=0.62}{%
     \begin{tikzcd}
        \lstick{$q_0$} & \targ{} & \ctrl{1} & \qw \\
        \lstick{$q_1$} & \ctrl{-1} & \targ{} & \qw \\
    \end{tikzcd}
    }     

    \\
\hline 
 Depth & 7 & 6 & 2  \\
\Xhline{4\arrayrulewidth}
\caption{\label{tab:novel-identities} Comparing Quanto to other compilers for optimizing quantum circuits with unknown optimizations}
\end{longtable}}
\end{center}
\end{figure*}

\section{Discussion}
\subsection{Related Work}
\subsubsection{Automatically generating identities}
To the best of our knowledge, previous work on automatically discovering quantum circuit identities is limited to single-qubit gates \cite{2003quantumcircuitidentities}. In addition, these automatically generated identities have not been applied to optimizing quantum programs.

\subsubsection{Superoptimization}
Superoptimization is a compiler technique that was originally designed to find the optimal machine code for a sequence of instructions \cite{1987superoptimizer} \cite{2008superoptimizer} \cite{2006superoptimization}. Our approach is inspired by TASO \cite{2019taso}. TASO applies superoptimization to deep learning by optimizing tensor computation graphs. For a given set of tensor operators (e.g., addition, matrix convolution) TASO automatically finds identities and uses these as substitutions to optimize computation graphs of deep learning applications. There are some differences: In the TASO paper, they generate a graph and then go through the list of substitutions and try to see whether the substitution is a sub-graph in the original graph. They have a limited number of substitutions, in the thousands, so they can do that fairly efficiently. In our case, however, we have can have hundreds of thousands or millions of substitutions and the time complexity of that technique would grow large. Instead we split our circuit into sub-circuits, tiles, and then look for the hash of that tile in the hash table, which is a constant lookup. Therefore, our technique is more efficient for our case and only depends on the number of tiles in the circuit, which grows modestly.

\subsubsection{Quantum Optimization}
There are other compilers, \cite{2017qiskit, 2020tket, 2019voqc} but their focus is mapping the logical circuit to the quantum device. Therefore, they do not pay much attention to the potential logical transformations. This can be shown in our results -- other optimizers fail to recognize certain circuit identities. These optimizers have manual rules for logical substitutions and therefore are not finding the full range of optimization. 

\subsection{Limitations and Future Work}
One limitation of Quanto is the scalability of generating a large number of circuit identities, which limits the size (depth and number of qubits) of the circuit identities. This issue could be mitigated by applying pruning techniques during the generation of substitutions to remove redundant substitutions or by using heuristics to produce substitutions. Another limitation is missing out on potential substitutions, due to invalid tiles in the quantum circuit as shown in Figure \ref{fig:invalid_tile}.

To use gates with parameters in our optimizer, we can set the parameter to a small set of specific values when generating identities. Since many parameters are used for qubit rotations, and certain angles of rotation are very common in quantum algorithms, this approach gives some reasonable coverage of potential optimizations involving parameters.  However, our optimizer would be improved if it included symbolic identities, where substitutions for parametrized gates are generated regardless of the value of the parameter.

\subsection{Conclusion}
Quanto is the first quantum optimizer that automatically generates general circuit identities. Quanto finds novel optimizations and finds better optimizations for quantum circuits, compared to other compilers. This work contributes to reducing the circuit depth of quantum circuits, which could lead to less noise and more accurate execution of quantum algorithms.

\section{Methods}
\subsection{Circuit Identity Generator}
This section describes the Quanto circuit identity generator, which automatically generates circuit identities given a gate set, as well as the number of qubits and depth bounds. This generation algorithm finds all circuit identities given the number of qubits, circuit depth, and gate set. \subsubsection{Generation Algorithm}
Table \ref{tab:generation_inputs} shows the inputs to the generation algorithm.

\begin{table}
\begin{tabular}{ |p{0.1\textwidth} | >{\raggedright}p{0.15\textwidth} | >{\raggedright\arraybackslash}p{0.20\textwidth}| }
\hline
 \thead{Symbol} & \thead{Definition} & \thead{Description} \\
\hline
\textit{n} & number of qubits & maximum number of qubits of the circuit identities generated \\
\hline
\textit{d} & depth of the circuit & the maximum depth of the circuit identities generated \\
\hline
\textit{dp} & decimal point accuracy & the decimal points of the  floating-point  numbers  in  the  matrices  of  the gates \\
\hline
\textit{gs} & gate set & a list of gates to use in the circuit identities, including the identity gate
\\
\hline
\end{tabular}
\caption{\label{tab:generation_inputs}Inputs of the generation algorithm}
\end{table}

\textbf{Step 1: Represent the quantum circuit as a grid.}
We represent a quantum circuit as a grid, where the number of rows $m$ is the number of layers in the quantum circuit, also known as the circuit depth. The number of columns $n$ is the number of qubits in the quantum circuit. If there is no gate present at a particular row for a particular column, we place the Identity gate in that position.

\textbf{Step 2: For a single layer (depth = 1), generate all the possible circuits of quantum circuits using the gate set}.

\textit{a) Single-qubit gates only} 

If there are only single-qubit gates in the gate set, then each gate is applied to a single qubit. If $n$ is the number of qubits and $g$ is the number of single-qubit gates, then there are $g^n$ possible circuits.

\textit{Example:} Gate Set = [\textit{I, X, H}] and $n = 2$
where 
$$
 I = \begin{pmatrix}
    1 & 0 \\ 
    0 & 1
\end{pmatrix},
 X = \begin{pmatrix}
    0 & 1 \\ 
    1 & 0
\end{pmatrix},
H = \begin{pmatrix}
    \frac{1}{\sqrt{2}} & \frac{1}{\sqrt{2}} \\ 
    \frac{1}{\sqrt{2}} & -\frac{1}{\sqrt{2}}
\end{pmatrix}
$$

Table \ref{tab:single-qubit-possible circuits} shows all the possible circuits that would be generated for the gate set $[I, X, H]$

\begin{table}
{\renewcommand{\arraystretch}{1.5}%
\begin{tabular}{ |c | c | c| }
\hline
& &
\\

\adjustbox{scale=1}{%
 \begin{tikzcd}
    \lstick{$q_0$} & \gate{I} & \qw \\
    \lstick{$q_1$} & \gate{I} & \qw \\
\end{tikzcd}
} 

&

\adjustbox{scale=1}{%
 \begin{tikzcd}
    \lstick{$q_0$} & \circuitX & \qw \\
    \lstick{$q_1$} & \circuitX & \qw \\
\end{tikzcd}
} 

&

\adjustbox{scale=1}{%
 \begin{tikzcd}
    \lstick{$q_0$} & \circuitH & \qw \\
    \lstick{$q_1$} & \circuitH & \qw \\
\end{tikzcd}
} 

\\ 
\hline
[\textit{I, I}] & [\textit{X, X}] & [\textit{H, H}] 
\\
\hline
& &
\\
\adjustbox{scale=1}{%
 \begin{tikzcd}
    \lstick{$q_0$} & \gate{I} & \qw \\
    \lstick{$q_1$} & \circuitX & \qw \\
\end{tikzcd}
} 

&

\adjustbox{scale=1}{%
 \begin{tikzcd}
    \lstick{$q_0$} & \gate{I} & \qw \\
    \lstick{$q_1$} & \circuitH & \qw \\
\end{tikzcd}
} 

&

\adjustbox{scale=1}{%
 \begin{tikzcd}
    \lstick{$q_0$} & \circuitX & \qw \\
    \lstick{$q_1$} & \gate{I} & \qw \\
\end{tikzcd}
} 

\\ 
\hline
[\textit{I, X}] & [\textit{I, H}] & [\textit{X, I}] 
\\
\hline
& &
\\
\adjustbox{scale=1}{%
 \begin{tikzcd}
    \lstick{$q_0$} & \circuitX & \qw \\
    \lstick{$q_1$} & \circuitH & \qw \\
\end{tikzcd}
} 

&

\adjustbox{scale=1}{%
 \begin{tikzcd}
    \lstick{$q_0$} & \circuitH & \qw \\
    \lstick{$q_1$} & \gate{I} & \qw \\
\end{tikzcd}
} 

&

\adjustbox{scale=1}{%
 \begin{tikzcd}
    \lstick{$q_0$} & \circuitH & \qw \\
    \lstick{$q_1$} & \circuitX & \qw \\
\end{tikzcd}
} 

\\ 
\hline
[\textit{X, H}] & [\textit{H, I}] & [\textit{H, X}] 
\\
\hline
\end{tabular}}
\caption{\label{tab:single-qubit-possible circuits}possible circuits for single-qubit gates given gate set $[I, X, H]$}
\end{table}. In the case of the possible circuit $[H, X]$ the $H$ gate is applied to the first qubit and the $X$ gate is applied to the second qubit as shown in Table \ref{tab:single-qubit-possible circuits}.

\textit{b) Two-qubit and single-qubit gates} 

If there are two-qubit gates, then we split the two-qubit gate into the first and second qubits it operates on.

\textit{Example:} Gate Set = [\textit{I, CX}] and $n=3$ 
where

$$
CX = \begin{pmatrix}
    1 & 0 & 0 & 0\\ 
    0 & 1 & 0 & 0\\
    0 & 0 & 0 & 1\\
    0 & 0 & 1 & 0
\end{pmatrix}
$$

In the notation below, \textit{CXC} denotes applying the \textit{CX} gate to the $control$ qubit and \textit{CXT} denotes applying the \textit{CX} gate to the $target$ qubit.

\begin{table}
{\renewcommand{\arraystretch}{1.5}%
\begin{tabular}{ |c | c | c | c |}
\hline
& & &
\\

\adjustbox{scale=0.75}{%
 \begin{tikzcd}
    \lstick{$q_0$} & \gate{I} & \qw \\
    \lstick{$q_1$} & \gate{I} & \qw \\
    \lstick{$q_2$} & \gate{I} & \qw \\
\end{tikzcd}
} 

&

\adjustbox{scale=0.75}{%
 \begin{tikzcd}
    \lstick{$q_0$} & \ctrl{1} & \qw \\
    \lstick{$q_1$} & \targ{} & \qw \\
    \lstick{$q_2$} & \gate{I} & \qw \\
\end{tikzcd}
} 

&

\adjustbox{scale=0.75}{%
 \begin{tikzcd}
    \lstick{$q_0$} & \ctrl{2} & \qw \\
    \lstick{$q_1$} & \gate{I} & \qw \\
    \lstick{$q_2$} & \targ{} & \qw \\
\end{tikzcd}
} 

&

\adjustbox{scale=0.75}{%
 \begin{tikzcd}
    \lstick{$q_0$} & \targ{} & \qw \\
    \lstick{$q_1$} & \ctrl{-1} & \qw \\
    \lstick{$q_2$} & \gate{I} & \qw \\
\end{tikzcd}
} 

\\ 
\hline
[\textit{I, I, I}] & [\textit{CXC, CXT, I}] & [\textit{CXC, I, CXT}] & [\textit{CXT, CXC, I}] 
\\
\hline
\end{tabular}}

{\renewcommand{\arraystretch}{1.5}%
\begin{tabular}{ |c | c | c |}

\hline
& &
\\

\adjustbox{scale=1}{%
 \begin{tikzcd}
    \lstick{$q_0$} & \gate{I} & \qw \\
    \lstick{$q_1$} & \ctrl{1} & \qw \\
    \lstick{$q_2$} & \targ{} & \qw \\
\end{tikzcd}
} 

&

\adjustbox{scale=1}{%
 \begin{tikzcd}
    \lstick{$q_0$} & \targ{} & \qw \\
    \lstick{$q_1$} & \gate{I} & \qw \\
    \lstick{$q_2$} & \ctrl{-2} & \qw \\
\end{tikzcd}
} 

&

\adjustbox{scale=1}{%
 \begin{tikzcd}
    \lstick{$q_0$} & \gate{I} & \qw \\
    \lstick{$q_1$} & \targ{} & \qw \\
    \lstick{$q_2$} & \ctrl{-1} & \qw \\
\end{tikzcd}
} 

\\ 
\hline
[\textit{I, CXC, CXT}] & [\textit{CXT, I, CXC}] & [\textit{I, CXT, CXC}] 
\\
\hline
\end{tabular}}
\caption{\label{tab:two-qubit-possible circuits}Possible circuits for two-qubit gates given gate set $[I, CX]$}
\end{table}

Table \ref{tab:two-qubit-possible circuits} shows all the possible circuits that would be generated for the gate set $[I, CX]$. In the case of the possible circuit $[CXC, CXT, I]$ the $CX$ gate is applied to the first and second qubit where the first qubit is the $control$ qubit and the second qubit is the $target$ qubit. The $I$ gate is applied to the third qubit. 

\textit{Verifying the validity of the two-qubit gates}

$CXC$ and $CXT$ must both be present in the possible circuits; it is not possible to have one present without the other. For example, $[CXC, I, I]$ would be an invalid configuration. Quanto verifies that $CXC$ and $CXT$ are present in each possible circuit.

\textbf{Step 3: For multiple layers (depth $> 1$), generate all the possible circuits of the quantum circuits using the gate set.} For an additional layer in the circuit, we add all the possible circuits for a single layer onto the previous layer. 

\textit{Example}: For the gate set presented in Table \ref{tab:single-qubit-possible circuits}, we take each possible circuit and append it to the end of each possible circuit. Table \ref{tab:multiple-layer-possible circuits} illustrates appending the possible circuit $[X, X]$ to each possible circuit. This is repeated for $[H, H], [I, X], etc...$

\begin{table}
{\renewcommand{\arraystretch}{1.5}%
\begin{tabular}{ |c | c | c| }
\hline
& &
\\

\adjustbox{scale=1}{%
 \begin{tikzcd}
    \lstick{$q_0$} & \gate{I} & \circuitX & \qw \\
    \lstick{$q_1$} & \gate{I} & \circuitX & \qw \\
\end{tikzcd}
} 

&

\adjustbox{scale=1}{%
 \begin{tikzcd}
    \lstick{$q_0$} & \circuitX & \circuitX & \qw \\
    \lstick{$q_1$} & \circuitX & \circuitX & \qw \\
\end{tikzcd}
} 

&

\adjustbox{scale=1}{%
 \begin{tikzcd}
    \lstick{$q_0$} & \circuitH & \circuitX & \qw \\
    \lstick{$q_1$} & \circuitH & \circuitX & \qw \\
\end{tikzcd}
} 

\\ 
\hline
$[I, I], [X, X]$ & $[X, X], [X, X]$ & $[H, H], [X, X]$
\\
\hline
& &
\\
\adjustbox{scale=1}{%
 \begin{tikzcd}
    \lstick{$q_0$} & \gate{I} & \circuitX & \qw \\
    \lstick{$q_1$} & \circuitX & \circuitX & \qw \\
\end{tikzcd}
} 

&

\adjustbox{scale=1}{%
 \begin{tikzcd}
    \lstick{$q_0$} & \gate{I} & \circuitX & \qw \\
    \lstick{$q_1$} & \circuitH & \circuitX & \qw \\
\end{tikzcd}
} 

&

\adjustbox{scale=1}{%
 \begin{tikzcd}
    \lstick{$q_0$} & \circuitX & \circuitX & \qw \\
    \lstick{$q_1$} & \gate{I} & \circuitX & \qw \\
\end{tikzcd}
} 

\\ 
\hline
$[I, X], [X, X]$ & $[I, H], [X, X]$ & $[X, I], [X, X]$
\\
\hline
& &
\\
\adjustbox{scale=1}{%
 \begin{tikzcd}
    \lstick{$q_0$} & \circuitX & \circuitX & \qw \\
    \lstick{$q_1$} & \circuitH & \circuitX & \qw \\
\end{tikzcd}
} 

&

\adjustbox{scale=1}{%
 \begin{tikzcd}
    \lstick{$q_0$} & \circuitH & \circuitX & \qw \\
    \lstick{$q_1$} & \gate{I} & \circuitX & \qw \\
\end{tikzcd}
} 

&

\adjustbox{scale=1}{%
 \begin{tikzcd}
    \lstick{$q_0$} & \circuitH & \circuitX & \qw \\
    \lstick{$q_1$} & \circuitX & \circuitX & \qw \\
\end{tikzcd}
} 

\\ 
\hline
$[X, H], [X, X]$ & $[H, I], [X, X]$ & $[H, X], [X, X]$
\\
\hline
\end{tabular}}
\caption{\label{tab:multiple-layer-possible circuits}Possible circuits for multiple layers}
\end{table}

\textbf{Step 4: Calculate the unitary matrices and generate their fingerprints.} For each possible circuit, we find the \textit{unitary matrix} of the possible circuit. We adopt an idea from compiler superoptimization \cite{2006superoptimization} and compute a {\em fingerprint} for each possible circuit, which is a hash of the unitary matrix. Figure \ref{fig:output_matrix} shows a possible circuit and how to calculate the unitary matrix. The matrices in part (c) correspond to the gates in part (b) and (a) of Figure \ref{fig:output_matrix}. 

\begin{figure}[H]
(a) $$\adjustbox{scale=1}{%
 \begin{tikzcd}
    \lstick{$q_0$} & \circuitH & \circuitX & \qw \\
    \lstick{$q_1$} & \circuitX & \circuitX & \qw \\
\end{tikzcd}
}$$
(b) $$ (H \otimes X) \times (X \otimes X) $$

(c) \begin{equation*}
\resizebox{1\hsize}{!}{$%
\Bigg(\frac{1}{\sqrt{2}}
\begin{pmatrix} 
1 & 1 \\
1 & -1 
\end{pmatrix} \otimes 
\begin{pmatrix}
0 & 1 \\
1 & 0 
\end{pmatrix}
\Bigg)
\times
\Bigg(
\begin{pmatrix}
0 & 1 \\
1 & 0
\end{pmatrix} \otimes
\begin{pmatrix}
0 & 1 \\
1 & 0
\end{pmatrix} \Bigg) \\
= \frac{1}{\sqrt{2}} \begin{pmatrix}
1 & 0 & 1 & 0 \\
0 & 1 & 0 & 1 \\
-1 & 0 & 1 & 0\\
0 & -1 & 0 & 1
\end{pmatrix}
$}
\end{equation*}

\caption{(a) A possible circuit generated from a gate set (b) the gate representation of the circuit (c) the matrix calculation of the circuit to obtain the unitary matrix}
\label{fig:output_matrix}
\end{figure}

\begin{figure}
\begin{flushleft}
 (a)
\end{flushleft} $$
\adjustbox{scale=1}{%
 \begin{tikzcd}
    \lstick{$q_0$} & \circuitH & \circuitX & \ctrl{1} & \qw \\
    \lstick{$q_1$} & \gate{I} & \circuitX & \targ{} & \qw \\
\end{tikzcd}
} 
$$

\begin{flushleft}
 (b)
\end{flushleft}
$$\includegraphics[width=0.5\columnwidth]{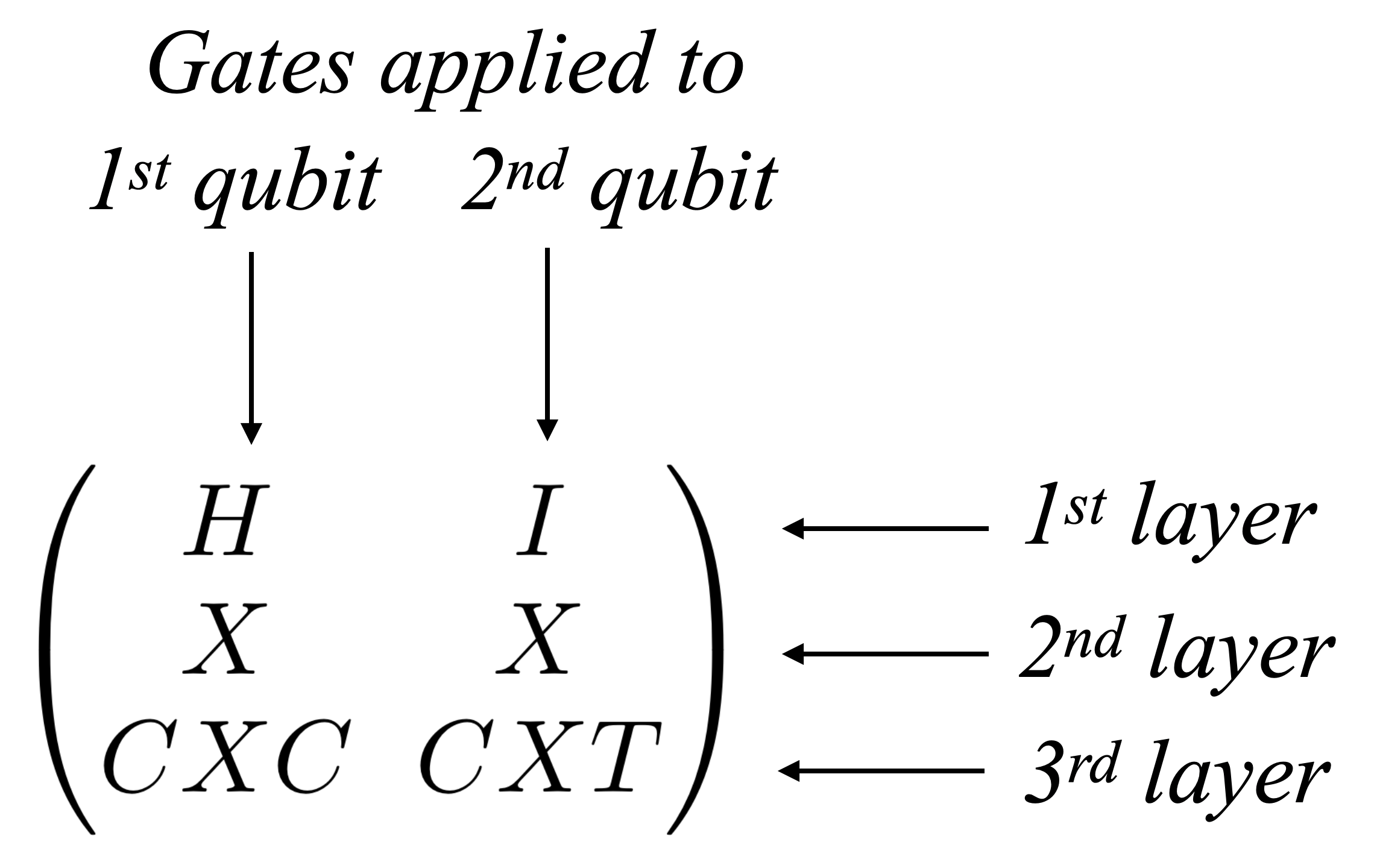}$$

\begin{flushleft}
 (c)
\end{flushleft} $$[[H, I], [X, X], [CXC, CXT]]$$

\caption{(a) Quantum circuit which can be represented as (b) a matrix, which can be represented computationally as (c)}
\label{fig:circuit_representation}
\end{figure}

\begin{figure}
\includegraphics[width=1\columnwidth]{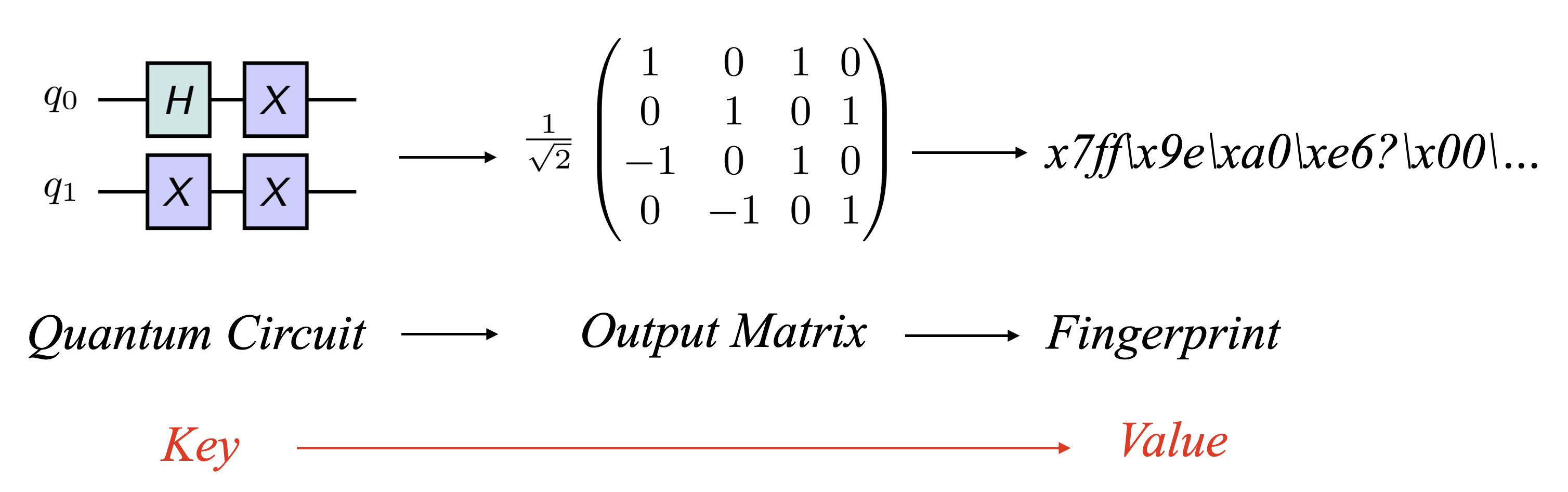}
\caption{A quantum circuit can be represented by an unitary matrix. The hash of this unitary matrix is its fingerprint.}
\label{fig:quantum_circuit_to_fingerprint}
\end{figure}

\textbf{Step 5: Create hash table of fingerprints.}
The hash table of fingerprints includes the quantum circuit as the key and the fingerprint of the quantum circuit as the value. This allows us to look up a quantum circuit's fingerprint in constant time during the optimizer stage, without having to recalculate the unitary matrix and the fingerprint of the quantum circuit. Figure \ref{fig:quantum_circuit_to_fingerprint} shows the sequence of steps to generate a hash table of fingerprints for each possible circuit. Table \ref{tab:hash-table-circuit-to-fingerprint} is an example hash table with one entry which shows the quantum circuit representation as the key and its corresponding fingerprint as the value.

\begin{table}
{\renewcommand{\arraystretch}{1.5}%
\begin{tabular}{ |c | c | }
\hline
\thead{Key} & \thead{Value} \\
\hline
$[[H, X], [X, X]]$ & $x7ffx9exa0xe6?x00…$ \\
\hline
\end{tabular}}
\caption{\label{tab:hash-table-circuit-to-fingerprint}Example of a hash table with one entry demonstrating the quantum circuit representation as the key and the fingerprint as the value. Note that the fingerprint in this table is only for illustrative purposes (as the actual fingerprint is too long to be displayed).}
\end{table}

\textbf{Step 6: Create hash table of circuit identities.}
The hash table of substitutions includes the fingerprints as keys and an array of matching circuit identities as values. For each possible circuit we generate, we check if its fingerprint is in the hash table of circuit identities. If it is, we add the possible circuit to the matching fingerprint. If it is not, we add the fingerprint to the hash table and add the possible circuit to the array of matching circuit identities. The identities are stored implicitly in the hashtable - any set of matrices associated with the same key are equal and can be substituted for each other. The identities are up to some (configurable) floating point precision. Table \ref{tab:hash-table-fingerprint-to-subs} is an example hash table with one entry which shows the fingerprint as the key and all the circuit identities that match to the fingerprint as the value. 

\begin{table}
{\renewcommand{\arraystretch}{1.5}%
\begin{tabular}{ |c | c | }
\hline
\thead{Key} & \thead{Value} \\
\hline
 $x7ffx9exa0xe6?x00…$ & \makecell{$[[X, X], [X, X]]$ \\
$[[H, H], [H, H]]$}
\\
\hline
\end{tabular}}
\caption{\label{tab:hash-table-fingerprint-to-subs}Hash table with one entry demonstrating the fingerprint as the key and the array of substitutions as the value.}
\end{table}

\subsection{Scaling}
To estimate the computational resources to generate circuit identities, we derived a formula for the total number of possible circuits to be generated. The symbols are defined in Table \ref{tab:scaling}.

\begin{equation}
     S_{l} =\sum_{r=0}^{\lfloor n/2 \rfloor}\frac{n!}{r! (n-2r)!} g^{n-2r} t^r
\end{equation}

\begin{equation}
    S = S_{l}^{d}
\end{equation}

\begin{table}
\centering
{\renewcommand{\arraystretch}{1.5}%
\begin{tabular}{ | c | l | }
\hline
\thead{Symbol} & \thead{Definition} \\
\hline
$S$ & total number of possible circuits \\
\hline
$S_{l}$ & number of possible circuits per depth \\
\hline
$g$ & number of single-qubit gates \\
\hline
$n$ & number of qubits \\
\hline
$t$ & number of two-qubit gates \\
\hline
$S$ & number of possible circuits for the depth \\
\hline
$d$ & depth \\
\hline
\end{tabular}}
\caption{\label{tab:scaling}Symbols for scaling formula}
\end{table}

As one can see, the number of possible circuits scales quickly with $n$ which is to be expected. We always use the optimizer, and if the size of the tile is equivalent to the size of the input circuit, there is only one tile, meaning we enumerated this circuit during generation.

\subsection{Optimizer}
This section describes the Quanto optimizer, which applies the automatically generated circuit identities to optimize a quantum circuit.
\subsubsection{Algorithm}
Table \ref{tab:optimizer-inputs} shows the inputs.

\begin{table}
\centering
{\renewcommand{\arraystretch}{1.5}%
\begin{tabular}{ |c | l | }
\hline
\thead{Symbol} & \thead{Definition} \\
\hline
quantum circuit & quantum circuit to optimize \\
\hline
$i$ & the maximum number of qubits of the tiles \\
\hline
$j$ & the maximum depth of the tiles \\
\hline
\end{tabular}}
\caption{\label{tab:optimizer-inputs}Possible circuits for multiple layers}
\end{table}

\textbf{Step 1: Generate tiles of the input quantum circuit.}
If the input circuit, which has depth (or width) $m$ and number of qubits (or length) $n$, is larger than the size of the circuit identities generated, we split a circuit into sub-circuits, called $tiles$. The tile width, $j$, is the maximum depth of the circuit identities and the tile length, $i$, is the maximum number of qubits of the circuit identities. We iterate through the rows and columns in the quantum circuit matrix and generate the indices of all possible tiles. The number of possible $i \times j$ tiles for an $n \times m$ circuit is $(n-i+1) \cdot (m-j+1)$.

\begin{figure}
\includegraphics[width=0.5\columnwidth]{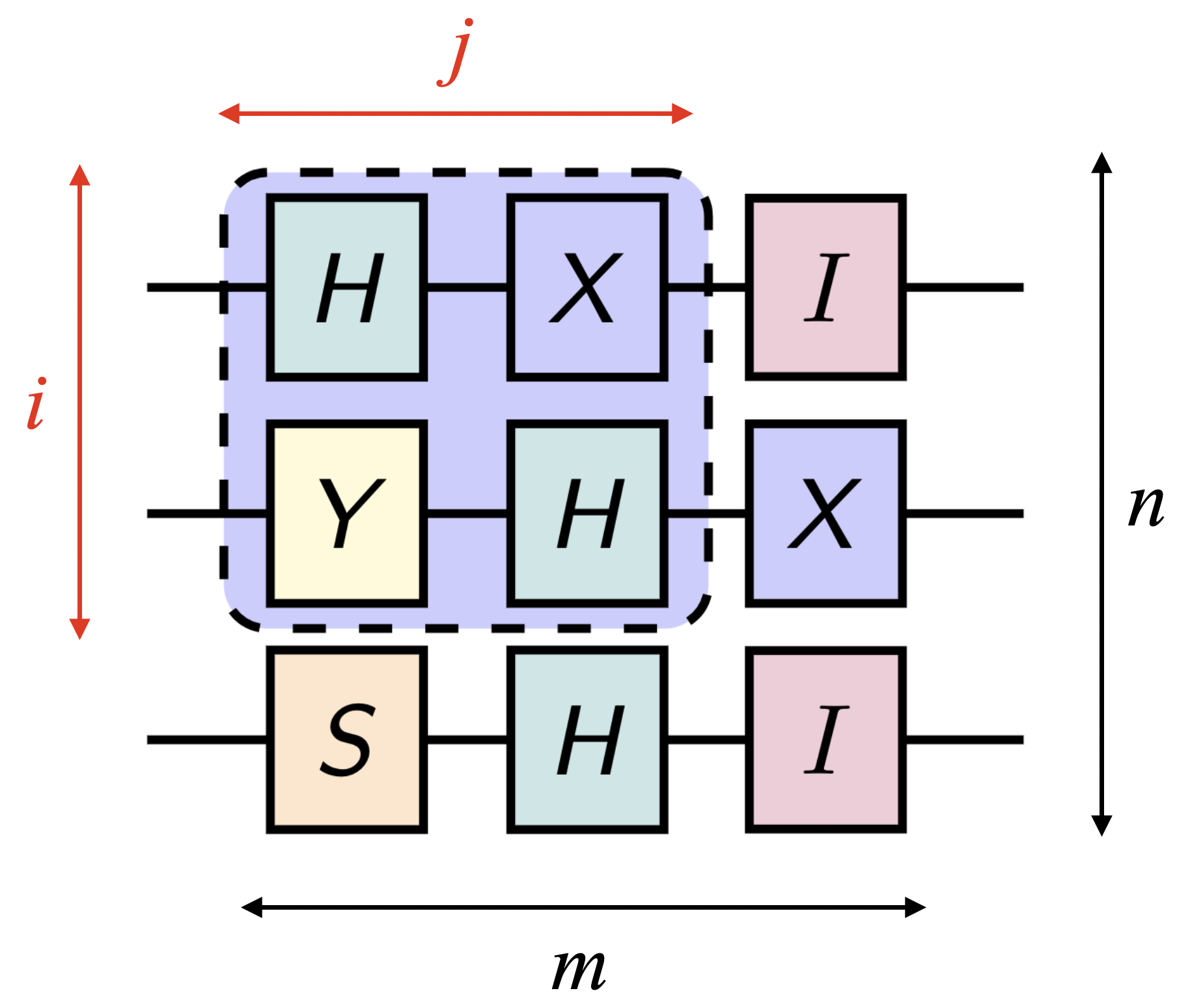}
\caption{\label{fig:tile_dimensions}Dimensions of a tile compared to the dimensions of a quantum circuit}
\end{figure}

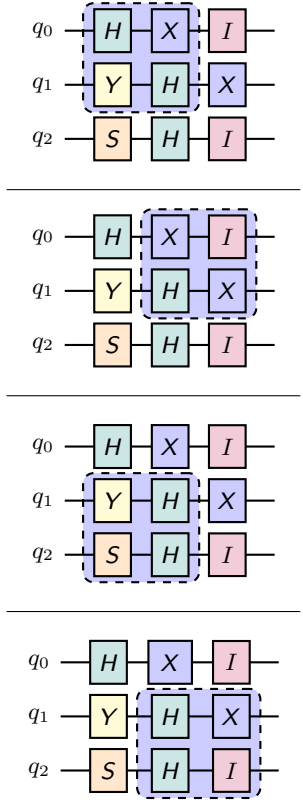
\begin{figure}

{\renewcommand{\arraystretch}{1}%
\begin{tabular}{ c }
\adjustbox{scale=1}{%
\begin{tikzcd}
    \lstick{$q_0$} & \circuitH\gategroup[2,steps=2,style={dashed,
    rounded corners,fill=blue!20, inner sep=4pt},
    background]{{\sc}} & \circuitX & \gate{I} & \qw \\
    \lstick{$q_1$} & \circuitY & \circuitH & \circuitX & \qw \\
    \lstick{$q_2$} & \circuitS & \circuitH & \gate{I} & \qw
\end{tikzcd}
} 

\\ \\
\hline
\adjustbox{scale=1}{%
 \begin{tikzcd}
    \lstick{$q_0$} & \circuitH & \circuitX\gategroup[2,steps=2,style={dashed,
    rounded corners,fill=blue!20, inner sep=4pt},
    background]{{\sc}} & \gate{I} & \qw \\
    \lstick{$q_1$} & \circuitY & \circuitH & \circuitX & \qw \\
    \lstick{$q_2$} & \circuitS & \circuitH & \gate{I} & \qw
\end{tikzcd}
} \\ \\
\hline
\\
\adjustbox{scale=1}{%
 \begin{tikzcd}
    \lstick{$q_0$} & \circuitH & \circuitX & \gate{I} & \qw \\
    \lstick{$q_1$} & \circuitY \gategroup[2,steps=2,style={dashed,
    rounded corners,fill=blue!20, inner sep=4pt},
    background]{{\sc}} & \circuitH & \circuitX & \qw \\
    \lstick{$q_2$} & \circuitS & \circuitH & \gate{I} & \qw
\end{tikzcd}
} \\ \\
\hline
\\
\adjustbox{scale=1}{%
 \begin{tikzcd}
    \lstick{$q_0$} & \circuitH & \circuitX\ & \gate{I} & \qw \\
    \lstick{$q_1$} & \circuitY & \circuitH \gategroup[2,steps=2,style={dashed,
    rounded corners,fill=blue!20, inner sep=4pt},
    background]{{\sc}} & \circuitX & \qw \\
    \lstick{$q_2$} & \circuitS & \circuitH & \gate{I} & \qw
\end{tikzcd}
} 
\end{tabular}}
\caption{\label{fig:tiles} All the possible tiles (shaded in blue) generated for a tile width and length = 2 for a quantum circuit with width and length = 3}
\end{figure}

Figure \ref{fig:tile_dimensions} shows the dimensions of the tiles and quantum circuit. Figure \ref{fig:tiles} shows all the possible tiles generated if we chose a tile with $j=2$ and $i=2$ in a quantum circuit with $m=3$ and $n=2$. Using the formula for the number of possible tiles, we see that there are four possible tiles, which corresponds with Figure \ref{fig:tiles}.

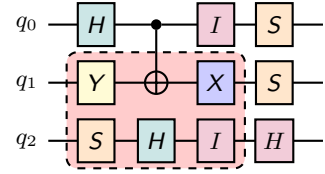
\begin{figure}
\adjustbox{scale=1}{%
 \begin{tikzcd}
    \lstick{$q_0$} & \circuitH & \ctrl{1} & \gate{I} & \circuitS & \qw \\
    \lstick{$q_1$} & \circuitY \gategroup[2,steps=3,style={dashed,
    rounded corners,fill=red!20, inner sep=4pt},
    background]{{\sc}}& \targ{}  & \circuitX & \circuitS & \qw \\
    \lstick{$q_2$} & \circuitS & \circuitH{} & \gate{I} & \gate{H} & \qw
\end{tikzcd}
} 
\caption{\label{fig:invalid_tile} An example of an invalid tile (shaded in red), because the $CX$ gate is cut off and is not first or last on its wire".}
\end{figure}

\begin{figure}
\adjustbox{scale=1}{%
 \begin{tikzcd}
    \lstick{$q_0$} & \circuitH & \ctrl{1} & \gate{I} & \circuitS & \qw \\
    \lstick{$q_1$} & \circuitY & \targ{} \gategroup[2,steps=3,style={dashed,
    rounded corners,fill=blue!20, inner sep=4pt},
    background]{{\sc}} & \circuitX & \circuitY & \qw \\
    \lstick{$q_2$} & \circuitS & \circuitH{} & \gate{I} & \gate{H} & \qw
\end{tikzcd}
} 
\caption{\label{fig:valid_tile} An example of a valid tile (shaded in blue), because even though the $CX$ gate is cut off, it is first or last on its wire.}
\end{figure}
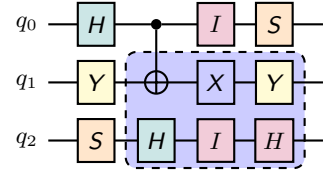

\textbf{Step 2: Check for valid tiles.}
Not all tiles generated from the circuit will be valid. If a two-qubit gate is not placed at the boundaries of the tile, then the tile is invalid if the two-qubit gate is cut off by the tile. In this case, we discard such tiles. We can keep a tile if the two-qubit gate is placed at the boundary of the tile even it has been cut off by the tile. Figure \ref{fig:invalid_tile} shows an example of an invalid tile, whereas Figure \ref{fig:valid_tile} shows an example of a valid tile. 

\textbf{Step 3: Search for tile in hash table of circuit identities.}
For each tile we have generated, we first search in the hash table of fingerprints.  If a tile contains a two-qubit gate that has been cut off \textit{only at the boundary} of the tile, then we first have to replace the two-qubit gate with an Identity gate in order to make the tile valid and searchable in the hash table of circuit identities. Part (a) in Figure \ref{fig:sub} shows this transformation. The now-valid tile is represented as a matrix, as shown in Part (b) in Figure \ref{fig:sub} and we can search for this matrix in the hash table of fingerprints. This gives us the fingerprint of the tile, without needing to calculate the unitary matrix of the tile. After we have obtained the fingerprint, we can search for circuit identities in the hash table of circuit identities as shown in Part (c) in Figure \ref{fig:sub}. Part (d) in Figure \ref{fig:sub} shows circuit identities for the tile.

Once we have the fingerprint, we then look up the possible substitutions by its fingerprint in the hash table of circuit identities.

\begin{figure}
a) $$\adjustbox{scale=0.8}{%
 \begin{tikzcd}
    \lstick{$q_0$} & \circuitH & \ctrl{1} & \gate{I} & \gate{I} & \qw \\
    \lstick{$q_1$} & \circuitY & \targ{} \gategroup[2,steps=3,style={dashed,
    rounded corners,fill=blue!20, inner sep=4pt},
    background]{{\sc}} & \circuitX & \circuitX & \qw \\
    \lstick{$q_2$} & \circuitS & \circuitY{} & \circuitH & \circuitH & \qw
\end{tikzcd}
}  \longrightarrow
\adjustbox{scale=0.8}{%
 \begin{tikzcd}
    \lstick{$q_0$} & \circuitH & \ctrl{1} & \gate{I} & \gate{I} & \qw \\
    \lstick{$q_1$} & \circuitY & \gate{I} \gategroup[2,steps=3,style={dashed,
    rounded corners,fill=blue!20, inner sep=4pt},
    background]{{\sc}} & \circuitX & \circuitX & \qw \\
    \lstick{$q_2$} & \circuitS & \circuitY{} & \circuitH & \circuitH & \qw
\end{tikzcd}
}$$
b) 
$$\begin{pmatrix}
I & Y \\
X & H \\
X & H
\end{pmatrix}
\longrightarrow \begin{tabular}{ |c | c | }
\hline
\thead{Key} & \thead{Value} \\
\hline
$[[I, Y], [X, H], [X, H]]$ & $x7ffx9exa0xe6?x00…$ \\
\hline
\end{tabular}
$$
c) 
$$\begin{tabular}{ |c | l | }
\hline
\thead{Key} & \thead{Value} \\
\hline
 $x7ffx9exa0xe6?x00…$ & \makecell{$[[I, Y], [I, H], [I, H]]$ \\
 $[[I, Y], [X, I], [X, I]]$ \\
$[[I, Y], [I, I], [I, I]]$}
\\
\hline
\end{tabular}$$

d) 
$$\adjustbox{scale=0.6}{%
 \begin{tikzcd}
    \lstick{$q_0$} & \circuitH & \ctrl{1} & \gate{I} & \gate{I} & \qw \\
    \lstick{$q_1$} & \circuitY & \gate{I} \gategroup[2,steps=3,style={dashed,
    rounded corners,fill=blue!20, inner sep=4pt},
    background]{{\sc}} & \gate{I} & \gate{I} & \qw \\
    \lstick{$q_2$} & \circuitS & \circuitY & \circuitH & \circuitH & \qw
\end{tikzcd}
}\adjustbox{scale=0.6}{%
 \begin{tikzcd}
    \lstick{$q_0$} & \circuitH & \ctrl{1} & \gate{I} & \gate{I} & \qw \\
    \lstick{$q_2$} & \circuitY & \gate{I} \gategroup[2,steps=3,style={dashed,
    rounded corners,fill=blue!20, inner sep=4pt},
    background]{{\sc}} & \circuitX & \circuitX & \qw \\
    \lstick{$q_1$} & \circuitS & \circuitY & \gate{I} & \gate{I} & \qw
\end{tikzcd}
} \adjustbox{scale=0.6}{%
 \begin{tikzcd}
    \lstick{$q_0$} & \circuitH & \ctrl{1} & \gate{I} & \gate{I} & \qw \\
    \lstick{$q_1$} & \circuitY & \gate{I} \gategroup[2,steps=3,style={dashed,
    rounded corners,fill=blue!20, inner sep=4pt},
    background]{{\sc}} & \gate{I} & \gate{I} & \qw \\
    \lstick{$q_2$} & \circuitS & \circuitY & \gate{I} & \gate{I} & \qw
\end{tikzcd}
}
$$
\caption{\label{fig:sub} (a) the invalid $CX$ gate in the tile is replaced with an Identity gate to make it a valid tile (b) the matrix representation of the tile is searched for in the hash table of fingerprints, which returns its fingerprint (c) the fingerprint is used to find the circuit identities for the tile in the hash table of circuit identities (d) the circuit identities for the tile}
\end{figure}

\textbf{Step 4: Apply substitutions via cost-based search.}
We then iterate through the circuit identities for each tile and evaluate the cost of each circuit identity. In this paper, we simply use the depth of the circuit (ignoring identity gates) as the circuit's cost; the cost function can be customized if desired. Table \ref{tab:cost} shows the cost for the circuit identities presented in Figure \ref{fig:sub}. In this case, circuit (iii) in Table \ref{tab:cost} has the lowest cost and will be used to substitute the tile in the quantum circuit.

\begin{table}
\begin{tabular}{ |l |c | c | }
\hline
\thead{} & \thead{Circuit Identity} & \thead{Cost} \\
\hline
& & \\
i &
\adjustbox{scale=1}{%
 \begin{tikzcd}
    \lstick{$q_0$} & \circuitH & \ctrl{1} & \gate{I} & \gate{I} & \qw \\
    \lstick{$q_1$} & \circuitY & \gate{I} \gategroup[2,steps=3,style={dashed,
    rounded corners,fill=blue!20, inner sep=4pt},
    background]{{\sc}} & \gate{I} & \gate{I} & \qw \\
    \lstick{$q_2$} & \circuitS & \circuitY & \circuitH & \circuitH & \qw
\end{tikzcd}
} & 3 \\ & & \\
\hline
& & \\
ii &
\adjustbox{scale=1}{%
 \begin{tikzcd}
    \lstick{$q_0$} & \circuitH & \ctrl{1} & \gate{I} & \gate{I} & \qw \\
    \lstick{$q_2$} & \circuitY & \gate{I} \gategroup[2,steps=3,style={dashed,
    rounded corners,fill=blue!20, inner sep=4pt},
    background]{{\sc}} & \circuitX & \circuitX & \qw \\
    \lstick{$q_1$} & \circuitS & \circuitY & \gate{I} & \gate{I} & \qw
\end{tikzcd}
} & 3 \\
& & \\
\hline
& & \\
iii &
\adjustbox{scale=1}{%
 \begin{tikzcd}
    \lstick{$q_0$} & \circuitH & \ctrl{1} & \gate{I} & \gate{I} & \qw \\
    \lstick{$q_1$} & \circuitY & \gate{I} \gategroup[2,steps=3,style={dashed,
    rounded corners,fill=blue!20, inner sep=4pt},
    background]{{\sc}} & \gate{I} & \gate{I} & \qw \\
    \lstick{$q_2$} & \circuitS & \circuitY & \gate{I} & \gate{I} & \qw
\end{tikzcd}
}  & 1 \\
& & \\
\hline
\end{tabular}
\caption{\label{tab:cost} Circuit identities and their cost}
\end{table}

\textbf{Step 5: Check for valid substitutions.}
Adding a substitution might make the entire quantum circuit invalid. For example, two-qubit gates may be cut off or not aligned properly. Additional steps, explained below, are needed to ensure the final circuit has valid two-qubit gates.

\textbf{Step 6: Apply the substitution to the circuit.}
We apply the substitution to the quantum circuit. If the original tile contained an invalid $CX$, then we have to do a couple more transformations before applying the substitution. Figure \ref{fig:apply-subs} shows the steps. Step (a) is equivalent to step (a) in Figure \ref{fig:sub}. In step (b), we apply the substitution with the lowest cost, which was the circuit identity (iii) in Table \ref{tab:cost}. We now need to transform this new circuit so that it is consistent with the gates in the original circuit. In particular, this new circuit has the $CX$ target gate replaced with an Identity gate. Step (b) replaces the Identity gate with the original $CX$ target gate. Step (c) shows the total transformation from the original circuit to an optimized circuit, after only one tile substitution. The Identity gates have been omitted to show that the depth of the circuit has decreased and therefore the circuit has been improved. We then continue to the next tile with the newly replaced substitution. 

\begin{figure}
\begin{flushleft}a)\end{flushleft} $$\adjustbox{scale=0.8}{%
 \begin{tikzcd}
    \lstick{$q_0$} & \circuitH & \ctrl{1} & \gate{I} & \gate{I} & \qw \\
    \lstick{$q_1$} & \circuitY & \targ{} \gategroup[2,steps=3,style={dashed,
    rounded corners,fill=blue!20, inner sep=4pt},
    background]{{\sc}} & \circuitX & \circuitX & \qw \\
    \lstick{$q_2$} & \circuitS & \circuitY{} & \circuitH & \circuitH & \qw
\end{tikzcd}
}  \longrightarrow
\adjustbox{scale=0.8}{%
 \begin{tikzcd}
    \lstick{$q_0$} & \circuitH & \ctrl{1} & \gate{I} & \gate{I} & \qw \\
    \lstick{$q_1$} & \circuitY & \gate{I} \gategroup[2,steps=3,style={dashed,
    rounded corners,fill=blue!20, inner sep=4pt},
    background]{{\sc}} & \circuitX & \circuitX & \qw \\
    \lstick{$q_2$} & \circuitS & \circuitY{} & \circuitH & \circuitH & \qw
\end{tikzcd}
}$$

\begin{flushleft}b)\end{flushleft} $$\adjustbox{scale=0.8}{%
 \begin{tikzcd}
    \lstick{$q_0$} & \circuitH & \ctrl{1} & \gate{I} & \gate{I} & \qw \\
    \lstick{$q_1$} & \circuitY & \gate{I} \gategroup[2,steps=3,style={dashed,
    rounded corners,fill=blue!20, inner sep=4pt},
    background]{{\sc}} & \gate{I} & \gate{I} & \qw \\
    \lstick{$q_2$} & \circuitS & \circuitY{} & \gate{I} & \gate{I} & \qw
\end{tikzcd}
}  \longrightarrow
\adjustbox{scale=0.8}{%
 \begin{tikzcd}
    \lstick{$q_0$} & \circuitH & \ctrl{1} & \gate{I} & \gate{I} & \qw \\
    \lstick{$q_1$} & \circuitY & \targ{} \gategroup[2,steps=3,style={dashed,
    rounded corners,fill=blue!20, inner sep=4pt},
    background]{{\sc}} & \gate{I} & \gate{I} & \qw \\
    \lstick{$q_2$} & \circuitS & \circuitY{} & \gate{I} & \gate{I} & \qw
\end{tikzcd}
} $$

\begin{flushleft}c)\end{flushleft} $$\adjustbox{scale=0.8}{%
 \begin{tikzcd}
    \lstick{$q_0$} & \circuitH & \ctrl{1} & \gate{I} & \gate{I} & \qw \\
    \lstick{$q_1$} & \circuitY & \targ{} \gategroup[2,steps=3,style={dashed,
    rounded corners,fill=blue!20, inner sep=4pt},
    background]{{\sc}} & \circuitX & \circuitX & \qw \\
    \lstick{$q_2$} & \circuitS & \circuitY{} & \circuitH & \circuitH & \qw
\end{tikzcd}
}  \longrightarrow \adjustbox{scale=0.8}{%
 \begin{tikzcd}
    \lstick{$q_0$} & \circuitH & \ctrl{1} & \qw \\
    \lstick{$q_1$} & \circuitY & \targ{} & \qw \\
    \lstick{$q_2$} & \circuitS & \circuitY{} & \qw
\end{tikzcd}
}
$$
\caption{\label{fig:apply-subs} Applying the substitution}
\end{figure}

\textbf{Step 7: Repeat process.}
Steps 1 - 6 are repeated for a fixed number of times. In our implementation, ten times were sufficient.

\subsection{Implementation}

Quanto is implemented in Python and can be used as an API. In particular, Quanto has the following modules:

\subsubsection{Circuit module}
Quanto can take \textit{qasm} (quantum assembly language) files \cite{2017openqasm} as input, \cite{2006quantumdesigntools, 2005qasm2circ, 2005qasmtools, 2005quale, 2016squash}, which is a popular way to construct circuits and makes our optimizer compatible with other quantum circuit frameworks, such as Qiskit \cite{2017qiskit} and Cirq \cite{2018cirq}, which can input/output \textit{qasm} files. Our code converts the \textit{qasm} file into our object representation of a circuit as shown in Figure \ref{fig:circuit_representation}. We represent a circuit as a grid, where the rows represent the layers and the columns represent the gates applied to the qubits.

\subsubsection{Subs module}
The user can specify a gate set to generate circuit identities. Quanto also has several pre-set gate set options apart from customized gate sets. If no gate set is specified, then our code automatically detects all the gates used in the input circuit and uses that as the gate set. This module generates the circuit identities and builds the circuit identities database is used for the optimizer module.

\subsubsection{Optimize module}
The optimize module takes the generated circuit identities database and optimizes the input circuit. It outputs an optimized circuit, along with metrics, such as the final circuit depth. The circuit can be saved as a \textit{qasm} file or a qiskit circuit (applicable for certain gates).
\\
%

\FloatBarrier
\bibliographystyle{plain}
\bibliography{references}

\begin{thebibliography}{10}

\bibitem{2017qiskit}
Qiskit.
\newblock \url{https://qiskit.org/}, 2017.

\bibitem{2018cirq}
Cirq.
\newblock \url{https://quantumai.google/cirq}, 2018.

\bibitem{2005quale}
S.~Balensiefer, Lucas Kreger-Stickles, and M.~Oskin.
\newblock Quale: quantum architecture layout evaluator.
\newblock 2005.

\bibitem{2006superoptimization}
Sorav {Bansal} and Alex {Aiken}.
\newblock {Automatic Generation of Peephole Superoptimizers}.
\newblock 2006.

\bibitem{2008superoptimizer}
Sorav Bansal and Alex Aiken.
\newblock Binary translation using peephole superoptimizers.
\newblock In {\em Proceedings of the 8th USENIX Conference on Operating Systems
  Design and Implementation}, OSDI'08, page 177–192, USA, 2008. USENIX
  Association.

\bibitem{2005qasm2circ}
I.~Chuang.
\newblock qasm2circ.
\newblock \url{http://www.media.mit.edu/quanta/qasm2circ/}, 2005.

\bibitem{2021gates}
Gavin~E. Crooks.
\newblock Gates, states, and circuits.
\newblock \url{https://threeplusone.com/pubs/on_gates.pdf}, 2021.

\bibitem{2005qasmtools}
A.~Cross.
\newblock qasm-tools.
\newblock \url{http://www.media.mit.edu/quanta/quanta-web/projects}, 2005.

\bibitem{2017openqasm}
A.~Cross, L.~Bishop, J.~Smolin, and J.~Gambetta.
\newblock Open quantum assembly language.
\newblock {\em arXiv: Quantum Physics}, 2017.

\bibitem{2016squash}
M.~Dousti, A.~Shafaei, and Massoud Pedram.
\newblock Squash 2: a hierarchical scalable quantum mapper considering ancilla
  sharing.
\newblock {\em Quantum Inf. Comput.}, 16:332--356, 2016.

\bibitem{2011equivalentquantumcircuits}
Juan~Carlos {Garcia-Escartin} and Pedro {Chamorro-Posada}.
\newblock {Equivalent Quantum Circuits}.
\newblock {\em arXiv e-prints}, page arXiv:1110.2998, October 2011.

\bibitem{2019voqc}
Kesha Hietala, Robert Rand, Shih{-}Han Hung, Xiaodi Wu, and Michael Hicks.
\newblock A verified optimizer for quantum circuits.
\newblock {\em Proc. {ACM} Program. Lang.}, 5({POPL}):1--29, 2021.

\bibitem{2019taso}
Zhihao Jia, Oded Padon, James~J. Thomas, Todd Warszawski, M.~Zaharia, and
  A.~Aiken.
\newblock Taso: optimizing deep learning computation with automatic generation
  of graph substitutions.
\newblock {\em Proceedings of the 27th ACM Symposium on Operating Systems
  Principles}, 2019.

\bibitem{2003quantumcircuitidentities}
Chris {Lomont}.
\newblock {Quantum Circuit Identities}.
\newblock {\em arXiv e-prints}, pages quant--ph/0307111, July 2003.

\bibitem{1987superoptimizer}
Henry Massalin.
\newblock Superoptimizer: a look at the smallest program.
\newblock {\em {ACM} {SIGARCH} Computer Architecture News}, 15(5):122--126,
  November 1987.

\bibitem{Note1}
IBM recently changed their native gate set to \{CX, ID, RZ, SX, X\}.

\bibitem{2019nisq}
John {Preskill}.
\newblock {Quantum Computing in the NISQ era and beyond}.
\newblock 2019:R62.001, January 2019.

\bibitem{2020tket}
Seyon Sivarajah, Silas Dilkes, Alexander Cowtan, Will Edgington, and Ross
  Duncan.
\newblock Tket: A retargetable compiler for nisq devices.
\newblock 03 2020.

\bibitem{2006quantumdesigntools}
K.~Svore, A.~Aho, A.~Cross, I.~Chuang, and I.~Markov.
\newblock A layered software architecture for quantum computing design tools.
\newblock {\em Computer}, 39:74--83, 2006.

\end{thebibliography}

\end{document}